\documentclass[aps,prb,reprint,superscriptaddress]{revtex4-2}
\usepackage[utf8]{inputenc}
\usepackage{graphicx}
\usepackage{subfigure}
\usepackage{xcolor}

\newcommand{\bfk}{{\bf k}}

\newcommand{\bfq}{{\bf q}}
\newcommand{\bfr}{{\bf r}}

\newcommand{\be}{\begin{eqnarray}}
\newcommand{\ee}{\end{eqnarray}}

\newcommand{\wbe}{\begin{widetext}}
\newcommand{\wee}{\end{widetext}}

\usepackage{bbm}
\usepackage{amsmath,leftidx}
\usepackage{graphicx}
\usepackage{times}
\usepackage{CJK}
\usepackage{color,colortbl}
\usepackage[normalem]{ulem}

\begin{document}


\title{Two-dimensional Paired Topological Superfluids of Rydberg Fermi Gases}

\author{Ching-Yu Huang}
\affiliation{ Department of Applied Physics, Tunghai University, Taichung 40704,Taiwan}

\author{Jiapei Zhuang}
\affiliation{ Department of Physics, National Tsing Hua University, Hsinchu 30013, Taiwan}

\author{Po-Yao Chang}
\affiliation{ Department of Physics, National Tsing Hua University, Hsinchu 30013, Taiwan}

\author{Daw-Wei Wang}
\affiliation{Department of Physics, National Tsing Hua University, Hsinchu 30013, Taiwan}
\affiliation{Physics Division, National Center for Theoretical Sciences, Taipei 10617, Taiwan}
\affiliation{Frontier Center for Theory and Computation, National Tsing Hua University, Hsinchu 30013, Taiwan}
\affiliation{Center for Quantum Technology, National Tsing Hua University, Hsinchu 30013, Taiwan}

\begin{abstract}
We systematically investigate the topological properties of spin polarized Rydberg-dressed fermionic atoms loaded in a bilayer optical lattice. 
Through tuning the Rydberg coupling strength and the inter-layer tunneling amplitude, we identify different types of topological superfluid states generated from the inter-layer pairing and relative gauge phase modulation of the couples 2D $p$-wave superfluids. 
These phases includes gapped/gapless with/without time reversal symmetry. 
One of the most interesting states is a gapless paired topological superfluid with both the time-reversal symmetry and particle-hole symmetry.
This state is equivalent  to a topological Kondo lattice model
with the spin-orbit coupling, an in-plane magnetic field, and an additional
particle-hole symmetry~\cite{Kondo_Lattice}. 
The flexibility of experimental manipulation in such Rydberg-dressed ferminoic systems therefore becomes a promising system for realizing interesting topological superfluids. 
\end{abstract}

\date{\today}

\maketitle
\section{ Introduction}

A Majorana fermion is defined to be its own antiparticle~\cite{MF}, and is a hypothetical particle in the theoretical high energy physics. 
In condensed matter systems, a Majorana fermion can appear as a localized edge state, reflecting the topological feature in the bulk of system. 
It is known that systems with topological feature can be classified as the symmetry-protected topological (SPT) orders
, which  are robust against local perturbations for a given on-site symmetries~\cite{classification}.
Ground states of nontrivial SPT phases cannot be continuously connected to trivial product states without either closing the gap or breaking the protecting symmetry~\cite{Bosonic_SPT}. 
One of the mostly studied SPT phases is proposed by Kitaev~\cite{Kitaev_1D} for one-dimensional (1D) $p$-wave superconductor. 
The Majorana zero mode (MZM) of such system may be applied for quantum computation through braiding, and certain experimental signature has been proposed in an ordinary $s$-wave superconducting wire with strong spin-orbital coupling through proximity effect~\cite{Proximity_MF,S_Heterostructures,tunable_S_device,Spin_Singlet_SC,S_SC_Heterostructures, Signatures_of_MF,Zero_bias_peaks,ac_Josephson,Anomalous_Zero_Bias_Conductance} or even in systems of ultracold atoms~\cite{MZM_Fermionic_Cold_Atoms, MF_Cold_Atom_Quantum_Wires, 2D_Spin_Orbit, s_Wave_SF}. 
However non-ambiguous evidence is still lacking probably due to the poor signal-to-noise ratio for a localized MZM.
Recently, some extensions of Kitaev's 1D model by including inter-chain tunneling~\cite{multichains}, dimerization  \cite{dimerize}, and long-ranged pairing~\cite{long_range_hopping_and_pairing, long_range_interaction} have also been proposed.

To stabilize interesting phases (nematic states, classical crystals, or quantum solids) in condensed matter systems,
the interactions are required to be strong enough to compete with the Fermi energy. One have to search and synthesize strongly correlated materials, 
which is a difficult task. 
On the other hand, the Rydberg atoms provide a better playground for investigating interesting phases from its highly tunability. 
E.g., the length scale and strength of the effective inter-atom interaction can be manipulated easily by external fields \cite{Dudin,Raitzsch,Pritchard,Nipper,Schaus}. 
In addition to the blockade effect for on-resonant excitations \cite{Jaksch,Lukin,Urban,Gaetan}, one can also apply a far-detuned weak field  to generate an effective Rydberg-dressed interaction (RDI), which has a soft core and a finite interaction range (~\cite{Henkel1,Henkel2,Balewski}. Theoretical calculations show that a repulsive RDI in a Bose gas may lead to a supersolid droplet phase~\cite{boson_crystal,Cinti,Pupillo,Henkel1,Henkel2}, while an attractive RDI induces a 3D bright soliton~\cite{Maucher}. 
For a Rydberg Fermi gas, some topological phases are also predicted for an attractive~\cite{Xiong} or repulsive interaction in an optical lattice near half-filling~\cite{Li}. Recently, Rydberg-dressed effective interaction has been observed for two individually trapped atoms~\cite{Jau} and in a 2D optical lattice by measuring the spin correlation~\cite{Zeiher}.

\renewcommand{\arraystretch}{1.4}
 \begin{table}[htb]
	\centering  
 \begin{tabular}{ |c|c| c| c |   }
 \hline
$t_z=0$  &
$\phi_{\uparrow/\downarrow}= 0$  &
$\phi_{\uparrow/\downarrow}=\pi/2$ &
$\phi_{\uparrow/\downarrow}=0/\pi$  \\
 \hline
$\alpha_{\uparrow/\downarrow}=0$    
&  $\circ$  &   $\circ$ & $\circ$ \\ \hline
$\alpha_{\uparrow/\downarrow}=0/\pi$  
& $\times$  $^\star$  &   $\circ$ &  $\circ$\\
  \hline
\end{tabular} 
\\[0.25cm]
\renewcommand{\arraystretch}{1.2}
 \begin{tabular}{ |c|c| c| c |   }
 \hline
$t_z \neq 0$    &
$\phi_{\uparrow/\downarrow}= 0$  &
$\phi_{\uparrow/\downarrow}=\pi/2$ &
$\phi_{\uparrow/\downarrow}=0/\pi$  \\
 \hline
$\alpha_{\uparrow/\downarrow}=0$ 
&  $\circ$ &  $\circ$ & $\times$  \\  \hline
$\alpha_{\uparrow/\downarrow}=0/\pi$    
&  $\times$  & $\times$  & $\times$ $^{\#}$ \\ \hline
	\end{tabular}
\caption{ 
The 12 representative cases and their types classified by their parameters and associate energy gap. 
$\circ$ indicates gapped spectrum and $\times$ indicates the gapless spectrum.
$\star$ and $ \# $ are  two cases discussed in main text.  
The former one preserves both the time-reversal symmetry and the particle-hole symmetry [symmetry class DIII],
and the later one only  preserves the particle-hole symmetry [symmetry class D].
}
\label{table:symmetry}
\end{table}

In this paper, we systematically study the topological properties of the ground states of fermionic Rydberg atoms loaded in a bilayer optical lattice. The effective Rydberg-dressing interaction is tuned to be attractive in all directions and hence generates the in-plane pairing ($\Delta_x^\sigma$ and $\Delta^\sigma_y$) as well as the inter-layer pairing ($\Delta_z$) at the same time (here $\sigma=\uparrow,\downarrow$ stands for the upper/lower layer index). 
Similar to the emergent gapless phases induced by inversion symmetry breaking in three dimensions~\cite{Shuichi_Murakami_2007} 
, the transitions between two topological phases might go through gapless phases. 
In this work, we focus on the topological properties of gapless phases. 
We first find that the topological properties of the ground state also depend not only on the system parameters, but also depend on the relative gauges phases between these order parameters. 
By calculating the entanglement spectrum, topological charge of the gapless points, and eigenstate energy/wavefunctions of a finite size systems, 
we identify two types of interesting gapless phases as follows. 
\textbf{Type I}: the topological gapless phase with the particle-hole symmetry ($\mathcal{P}$), belonging to symmetry class D. 
\textbf{Type II}:  the topological gapless phase with both the time-reversal symmetry ($\mathcal{T}$) and the particle-hole symmetry ($\mathcal{P}$) belonging to symmetry class DIII. 
The Type I phase further breaks the $C_4$ in-plane rotational lattice symmetry, making different zero energy flat bands in $x$ and $y$ directions. 
We also find that Type II phase could be mapped onto a topological Kondo lattice models with spin-orbital coupling,
an external magnetic field, and an additional particle-hole symmetry~\cite{Kondo_Lattice}.
The paired topological superfluids studied here is therefore a quiet general and practical system for the investigation of many interesting topological properties.  These phases are summarized in Table~\ref{table:symmetry}.

The manuscript is organized as follows: 
In Sec.~\ref{sec:system} we briefly describe the physical system of Rydberg atoms in 2D bilayer optical lattice. 
We derive the effective mean-field Hamiltonian and discuss their symmetry properties.  
In Sec.~\ref{sec:method} we explain how to detect the topological properties using entanglement spectrum/entropy as well as finite size calculation. 
We then show our calculation results for the two types of phases in Sec.~\ref{sec:results_typeI} and Sec.~\ref{sec:results_typeII}. 
We then provide further discussion in Sec.~\ref{sec:discussion} and conclude our paper in Sec.~\ref{sec:conclusion}. 
In Appendix~\ref{sec:eigen-mode}, we explicitly calculate the single-particle energy spectra of mean-field Hamiltonian.

\section{Physical System, Model Hamiltonian, and Symmetry Properties}
\label{sec:system}

\subsection{Effective Interaction}

We consider a single-species Fermi gas, where each atom is weakly coupled to its Rydberg excited state by an off-resonant two photon transition via an intermediate state. In the far detuning and weak coupling limit, we can apply the standard perturbation and adiabatic approximation~\cite{Henkel1} and obtain the effective Rydberg-dressed interaction between dressed state atoms: 
$V_{\rm RD}(\mathbf{r})=\frac{U_{0}}{1+(r/R_{c})^{6}}$~\cite{Henkel2,Maucher,Honer,Xiong}, where $\Omega$ and $\Delta$ are the effective Raman coupling and detuning respectively. 
$U_{0}\equiv(\Omega/2\Delta)^{4} C_{6}/R_{c}^{6}$, 
and $R_{c}\equiv (C_{6}/2|\Delta|)^{1/6}$ are the interaction strength and the averaged soft-core radius (Rydberg blockade radius). $C_{6}$ is the averaged van der Waals coefficient, which can be shown to be negative ($U_0<0$) for all orbital states when exciting $^{6}Li$ to a $|nD\rangle$ state with $n>30$~\cite{Walker,Singer2}.
Note that the radius $R_c$ could be tuned by the external laser detuning and strength independently by keeping the total interaction amplitude ($U_0$) the same. The decay rate of the two-photon process can be strongly reduced by choosing larger detuning in the first transition, and the atomic loss rate due to the orbital level crossing in the short-distance can be also reduced by Pauli exclusion principle.  More details of possible experimental setting can be found in Ref.~\cite{Xiong}.

\subsection{Mean field Hamiltonian in real space}

The full system Hamiltonian for spin polarized Rydberg Fermi gas in a 2D bilayer optical lattice can be expressed as follows in the real-space coordinates:
\wbe
\be
H =
&& 
-\sum_{\bfr, \sigma=\uparrow,\downarrow}
   \Big[ \mu c^\dagger_{\bfr,\sigma}c^{}_{\bfr,\sigma }  
  +t   c^\dagger_{\bfr,\sigma}c^{}_{\bfr+\hat{x},\sigma}   
+t   c^\dagger_{\bfr,\sigma}c^{}_{\bfr+\hat{y},\sigma}  \Big]
 -t_z \sum_{\bfr}
    c^\dagger_{\bfr,\uparrow}c^{}_{\bfr,\downarrow} +  h.c.
\\
 +&&\frac{1}{2}  \sum_{\bfr,\bfr'} 
\Big[ 
   V_{\parallel } (\bfr-\bfr')  \left(    
    c^{\dagger}_{\bfr,\uparrow}c^{\dagger}_{\bfr',\uparrow}c^{}_{\bfr',\uparrow} c^{}_{\bfr,\uparrow}+
    c^{\dagger}_{\bfr,\downarrow}c^{\dagger}_{\bfr',\downarrow}c^{}_{\bfr',\downarrow} c^{}_{\bfr,\downarrow} \right) 
+ V_{\perp}( \bfr-\bfr') \left( 
   c^{\dagger}_{\bfr,\uparrow}c^{\dagger}_{\bfr',\downarrow}c^{}_{\bfr',\downarrow} c^{}_{\bfr,\uparrow} 
+   c^{\dagger}_{\bfr,\downarrow}c^{\dagger}_{\bfr',\uparrow}c^{}_{\bfr',\uparrow} c^{}_{\bfr,\downarrow}   \right) \Big]    \nonumber 
\label{H_original}
\ee
\wee
where $ c^\dagger_{\bfr}  (c^{}_{\bfr,\sigma}) $ is the creation (annihilation) operator of fermions for the layer index $\sigma$ and the in-plane coordinate $\bfr=(i_x, i_y)$.  
$\mu$, $t$ and $t_z$ are the chemical potential, the intra-layer hopping amplitude, and the inter-layer hopping amplitude.  
$V_{\perp }( \bfr-\bfr') $ and $V_{\parallel }( \bfr-\bfr')$  are the interlayer and intra-layer interaction, respectively.

In order to simplify the system for a concrete discussion and comparison, below we will consider the situation when the Rydberg blockade radius, $R_c$, is tuned to be close to the in-plane lattice spacing, $a$, and the interlayer distance, $d$, so that only the nearest neighboring attractive interaction are considered throughout this paper. Extension to longer range interaction could be also carried out easily, but we believe all the quantitative results we discuss below will not change. 

By considering the nearest-neighboring-site interaction only, we could simply obtain the following mean-field Hamiltonian by introducing the nearest neighboring coupling:
\be
H =&& \sum_{\bfr, \sigma=\uparrow,\downarrow} 
  \Big[ - \mu c^\dagger_{\bfr,\sigma}c^{}_{\bfr,\sigma }  
 -t   c^\dagger_{\bfr,\sigma}c^{}_{\bfr+\hat{x},\sigma}    
-t    c^\dagger_{\bfr,\sigma}c^{}_{\bfr+\hat{y},\sigma}   
\nonumber\\
&&
+  \Delta_{x}^ {\sigma }  c^{\dagger}_{\bfr,\sigma}c^{\dagger}_{\bfr+\hat{x},\sigma}  
+ i \Delta_{y}^{\sigma }  c^{\dagger}_{\bfr,\sigma}c^{\dagger}_{\bfr+\hat{y},\sigma} + h.c. \Big]
 \\ 
&& +\sum_{\bfr}
  \Big[ -t_z   c^\dagger_{\bfr,\uparrow}c^{}_{\bfr,\downarrow} 
  + \Delta_{z} c^{\dagger}_{\bfr,\uparrow}c^{\dagger}_{\bfr,\downarrow} + h.c. \Big] + \text{const}   \nonumber . 
\label{H_meanfield}.
\ee
Here $\Delta_{z}$ is the gap function between the upper ($\uparrow$) and bottom layer ($\downarrow$) on site $\bfr$, and 
$\Delta_{x(y)}^{\sigma}$ are the gap function between the nearest-neighboring sites for $\sigma=\uparrow,\downarrow$ layers. More precisely, we have
\be
\Delta_{x}^{\sigma}  
&&= e^{i \phi_{\sigma} } \Delta_p\equiv    V_{\parallel} (\hat{x})   \langle    c^{}_{\bfr+\hat{x},\sigma}c^{}_{\bfr,\sigma} \rangle  \notag
\\
\Delta_{y}^{\sigma}  &&=e^{i \alpha_{\sigma} }e^{i \phi_{\sigma} }  \Delta_p\equiv  V_{\parallel} (\hat{y})  \langle   c^{}_{\bfr+\hat{y},\sigma}c^{}_{\bfr,\sigma} \rangle,
\nonumber \\ 
 \Delta_{z}   
 &&=\Delta_s\equiv  V_{\perp}( \bfr-\bfr')   \langle   c^{}_{\bfr,\downarrow }c^{}_{\bfr, \uparrow} \rangle.
\ee
Here the four phases $\alpha_{\uparrow,\downarrow}$ and $\phi_{\uparrow,\downarrow}$ are all arbitrary by fixing the gauge phase of the inter-layer pairing, $\Delta_z\equiv \Delta_s$. We have set $\Delta_p$ and $\Delta_s$  are both real and positive.
The phase difference between $\Delta_{x }^{\sigma}$  and $\Delta_{y}^{\sigma}$  makes the system different from the ordinary $p_x \pm ip_y$ pairing system, and may generate interesting topological properties as we will study later. Throughout this paper, we set $\Delta_z=\Delta_s$ to be real and positive in order to fix the overall gauge phase.

\subsection{Mean-field Hamiltonian in momentum space}
\begin{figure}[htb]
\includegraphics[width=0.48\textwidth]{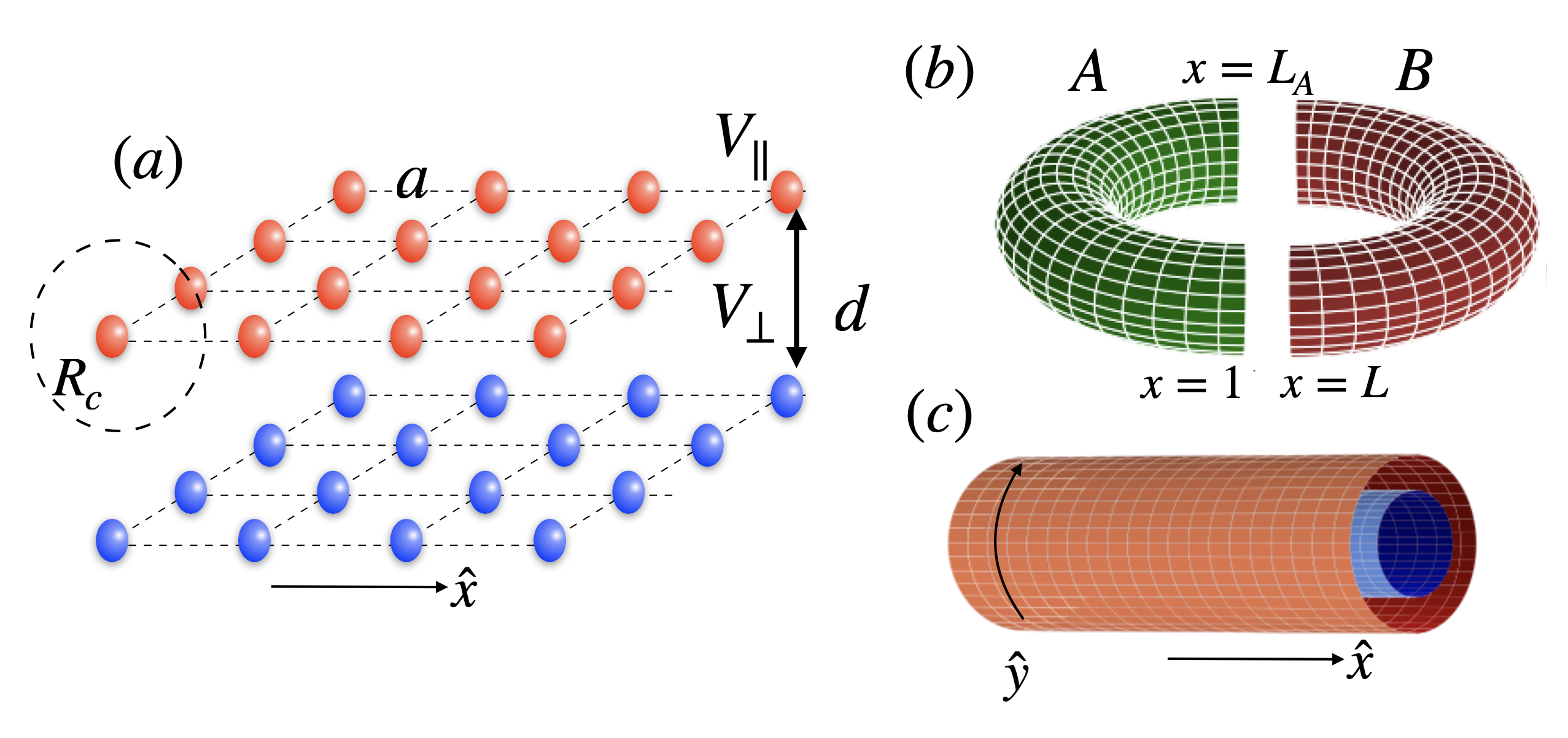}
\caption{ 
(a) The schematic figure for the system investigated in this paper: Rydberg-dressing spin-polarized fermions are loaded into a bilayer system with an in-plane square optical lattice. The Rydberg-dressing amplitude and radius, $R_c$, could be manipulated in a wide range of laser detuning and Rabi coupling strength. Herre we just consider the case when $R_c$ is similar to the in-plane lattice constant ($a$) and inter-layer spacing ($d$), so that only the nearest neighbouring interaction, $V_\|$, and the inter-layer on-site interaction, $V_\perp$, are included in the Hamiltonian for simplicity.
(b) The torus generated by the periodic boundary condition in both $x$ and $y$ direction of the bilayer system. When we consider the edge mode with an open boundary condition in $x$ axis, we cut the torus by keeping periodic boundary condition in the $y$ direction. As a result, the system becomes equivalent to two co-centered “finite cylinders” as shown in (c). After introducing the mean-field pairing and the Fourier transform along the $y$ direction, Hamiltonian becomes equivalent to a 4-leg ladder with the index $\pm k_y$ and $\sigma=\uparrow/\downarrow$, which could then be diagoanlized easily to get the full energy spectrum as well as the edge mode at the boundary of the $x$ axis. See the text for more details.
}   
\label{fig:structure}
\end{figure}

We analyze the mean-field Hamiltonian both for the periodic boundary condition and the open boundary condition. The former provides the band structure properties in the bulk while the later reveals the existence of possible topological protected boundary states.
To investigate the bulk properties with the periodic boundary condition, we transform the field operators as 
$c_{\bfr,\sigma}=  \frac{1}{N}\sum_{\bfk} c_{\bfk,\sigma}e^{i\bfr \bfk}$ with $\bfr= (i_x,i_y)$, $\bfk=(k_x, k_y )$ and $N$ being the total number of lattice sites in the 2D plane. The resulting Hamiltonian can be expressed to be (after neglecting a constant term) $H = \sum_{\bfk} \Psi_{\bfk}^{\dagger} H(\bfk)\Psi_{\bfk}^{}$ with 
\be
H (\bfk) 
\!\!= \!\!
\left( \begin{array}{cccc}
\varepsilon_{\bfk} & -t_z & (\tilde{\Delta}_{\bfk}^{\uparrow})^\dagger   &\Delta_s \\
-t_z &  \varepsilon_{\bfk}& -\Delta_s & ( \tilde{\Delta}_{\bfk}^{\downarrow} )^\dagger   \\
\tilde{\Delta}_{\bfk}^{\uparrow}  & -\Delta_s& -\varepsilon_{\bfk}  & t_z  \\
\Delta_s & \tilde{\Delta}_{\bfk}^{\downarrow}  &t_z  & -\varepsilon_{\bfk} 
\end{array}\right)
\!\!=\!\!
 \left( \begin{array}{cc}
H_0 &H_1^{\dagger}  \\
H_1& -H_0
\end{array}\right)
\label{H_original_2}
\ee
and the Nambu spinor 
$\Psi_{\bfk}  \equiv \left( c^{}_{\bfk,\uparrow},c^{}_{\bfk,\downarrow}, c^{\dagger}_{-\bfk,\uparrow},c^{\dagger}_{-\bfk,\downarrow}         \right )^T$. Here we have defined
\be
 &&\varepsilon_\bfk= -\mu  -2t\cos {k_x}  -2t \cos {k_y}, \nonumber \\
 && \tilde{\Delta}_{\bfk}^{\sigma} = 2i  e^{i \phi_{\sigma} } \Delta_p \; \big( \sin {k_x} + i e^{i \alpha_{\sigma} } \sin{k_y}  \big). 
\ee
and 
\be 
H_0&=&
\left( \begin{array}{cc}
\varepsilon_{\bfk} &-t_z \\
-t_z & \varepsilon_{\bfk} 
\end{array}\right)=\varepsilon_{\bfk } \sigma_0 - t_z \sigma_x
\\
H_1&=&
\left( \begin{array}{cc}
 \tilde{\Delta}_{\bfk}^{\uparrow} &-\Delta_s  \\
\Delta_s  &  \tilde{\Delta}_{\bfk}^{\downarrow}
\end{array}\right)=i\left( \Delta_s \sigma_0 +   \Delta_p \mathbf{d}_{\bfk} \cdot \vec{\sigma} \right)\times\sigma_y
\ee
where $\sigma_0=I$ and $\mathbf{ \sigma}_{\alpha}$ ($\alpha=x,y,z$) are Pauli matrices. 
The general form of the effective magnetic field, $\mathbf{d}_\bfk$, can be very complicated. Here we show three typical $\mathbf{d}_{\bfk}$ that
we will discuss in details in the following sections:
\begin{align}
&\mathbf{d}_{\bfk} = \left( 0, 2 \sin k_x+ 2 i \sin k_y, 0  \right), \quad (\alpha_{\uparrow/\downarrow},\phi_{\uparrow/\downarrow})=(0,0),&   \notag \\
&\mathbf{d}_{\bfk} = \left( 2 \sin k_y, 2  \sin k_x, 0  \right), \quad (\alpha_{\uparrow/\downarrow},\phi_{\uparrow/\downarrow})=(0,0/\pi),&  \notag \\
&\mathbf{d}_{\bfk} = \left( -2i\sin k_x, 2 i \sin k_y, 0  \right), \quad (\alpha_{\uparrow/\downarrow},\phi_{\uparrow/\downarrow})=(0/\pi,0/\pi).&
\end{align}

One should notice that the above Bogoliubov–de Gennes (BdG) Hamiltonian is similar to the noncentrosymmetric superconductors~\cite{PhysRevLett.105.097002,PhysRevLett.105.217001,PhysRevB.84.060504}. 
Due the in lack of inversion symmetry, odd and even pairings are mixed and the BdG Hamiltonian 
can be gapless.

\subsection{ The system symmetry }

Now we examine the symmetry class of the Hamiltonian by considering the time-reversal symmetry $(\mathcal{T})$, and the particle-hole symmetry $(\mathcal{P})$, chiral symmetry $(\mathcal{C})$ in the original BdG Hamiltonian. They are respectively defined as,
\be 
&& \mathcal{T}^{}  \; H( \bfk ) \; \mathcal{T}^{\dagger} = H^{*}(-  \bfk ) \nonumber\\ 
&& \mathcal{P}^{}  \; H( \bfk ) \; \mathcal{P}^{\dagger} = -H^{T}(-  \bfk ) \nonumber\\ 
&&  \mathcal{C}^{}  \; H( \bfk ) \; \mathcal{C}^{\dagger} = -H^{}(  \bfk ) \nonumber\\
\ee
Within the spinor representation, the first three symmetry operations could be also expressed by Pauli matrix in the isospin basis as following: 
\be 
&& \mathcal{T} = \sigma_0 \otimes i \sigma_y, \quad  \mathcal{T} ^2=-1 \nonumber\\ 
&& \mathcal{P} = \sigma_x \otimes \sigma_0, \quad    \mathcal{P}^2 = 1  \nonumber\\ 
&& \mathcal{C} =  \mathcal{P}\mathcal{T}^{\dagger} =  (\sigma_x\otimes \sigma_0) ( \sigma_0 \otimes -i\sigma_y) \nonumber\\
\ee

We note that unlike the chiral superconductor with $p_x+ip_y$ pairing which breaks the time-reversal symmetry,
the bilayer system we study here provides a possibility of the time-reversal symmetric state. 
More precisely, we find that for $t_z=0$, $\phi_{\uparrow}= -\phi_{\downarrow}$ and $ \alpha_{\uparrow}= \pi+ \alpha_{\downarrow}$,
the mean-field Hamiltonian is time-reversal invariant, $\mathcal{T}H(\bfk )\mathcal{T}^\dagger =  H^*(-\bfk )$.
In other words, the chiral directions of the two superfluids in the upper and lower layers are opposite, making the time-reversal symmetry restored if reversing the parity in the $z$ axis. Furthermore, we show that after a unitary transformation, the time-reversal symmetric case with $t_z=0$, $\phi_{\uparrow} =\phi_{\downarrow} =0$, $\alpha_{\uparrow}=0$, and $\alpha_{\downarrow}=\pi$ could be exactly mapped onto a  topological Kondo lattice model with spin-orbital coupling,
an external magnetic field, and an additional particle-hole symmetry ~\cite{Kondo_Lattice}.
On the other hand, our model is defined on the square lattice with the $C_4$ in-plane rotational lattice symmetry ($\mathcal{C}_4$).

In the rest of this paper, we always consider the cases when $\Delta_s$ and $\Delta_p$ are finite but with different combination of $t_z$, $\alpha_{\uparrow/\downarrow}$ and $\phi_{\uparrow/\downarrow}$.
More precisely, we consider the following $2\times 2\times 2=12$ representative system parameters with 2 values of $t_z$ ($=0$ or $\neq 0$), 2 values of $(\alpha_{\uparrow},\alpha_\downarrow)$ ($=(0,0)$ or $=(0,\pi)$), and 3 values of $(\phi_{\uparrow},\phi_\downarrow)$ ($=(0,0)$, $=(\pi/2,\pi/2)$ or $=(0,\pi)$). 
Moreover, we observe several gapless phases in our mean-field Hamiltonian.
In particular, we only show two cases describing gapless phases with two values of $t_z$ ($=0$ or $\neq 0$) by fixed $(\phi_{\uparrow},\phi_\downarrow)=(0,0)$ and  $(\alpha_{\uparrow},\alpha_\downarrow) =(0,\pi)$.

\section{  Topological properties  }
\label{sec:method}

\subsection{Topological index}

Topological quantum numbers are known to play a important role in characterizing the topological order phases.  
To compute the topological invariant that distinguishes these phases, we may consider a our 2D superconductor model described by a Hamiltonian $H(\bfk)$ in momentum space. 
By diagonalize $H(\bfk) |\psi_n(\bfk) \rangle = \varepsilon_n(\bfk)  |\psi_n(\bfk) \rangle  $, we can obtain a collection of bands $\varepsilon_n(\bfk)$ and  eigenstates  $\psi_n(\bfk)$ for integer $n$
The Chern number assigned to the $n$-th band based on Berry connection is defined by
\be
C_n = \frac{i}{2\pi}  \int d^2 \bfk [
\partial_{k_x} \langle \psi_n | \partial_{k_y} \psi_n \rangle  -  
\partial_{k_y} \langle \psi_n | \partial_{k_x} \psi_n \rangle    ]  
\ee

\subsection{ Entanglement spectrum}

Besides of the Chern number, a topological phase can be also characterized by the quantum entanglement between the subsystem and the environment~\cite{EN_TO_1,EN_TO_2, ES_FQHE,  ES_SPT_1,ES_SPT_2, Chang_2014}. 
Given a ground state wave function $|\Psi\rangle$, one can calculate the reduced density matrix, $\rho_A$, for a subsystem $A$ by tracing over the environment. 
The eigenvalues $\lambda_\alpha$ of the reduced density matrix is so-called ``entanglement spectrum"~\cite{ES_FQHE}, which carries nonlocal information and has been applied for calculating Berry phase and zero-energy edge states~\cite{berry_phase}.
For example, the degeneracy of the entanglement spectrum has recently been implemented to characterize the topological orders for some 2D quantum Hall states and for some 1D SPT phases~\cite{ES_FQHE}.

For 1D Kitaev model, $\lambda_\alpha$ is given by the eigenvalues of the block Green's function matrix, i.e.,  $G_{i,j}\equiv\langle c^{}_{i}c^\dagger_{j} \rangle$ 
with the site indices $i$ and $j$ inside the subsystem $A$. 
In Ref.~\cite{berry_phase, opes_1,opes_2}, it has been shown that the zero energy mode of the 1D Kitaev model corresponds to the degeneracy of $\lambda_{\alpha} =1/2$ in the entanglement spectrum, i.e. the pair of zero modes at the two ends of Kitaev's chain contribute the maximal entanglement between the subsystem $A$ and environment. 
Besides of entanglement spectrum, a topological phase transition can be also identified by the entanglement entropy of the subsystem (given by $S_A = - Tr \rho_A \log \rho_A$) after tracing out the environment.

It has been known that the entanglement spectrum $\lambda_{\alpha}$  of the subsystem $A$ can be obtained by diagonalizing the entire Green's function matrix $G_{ \bfr_1, \sigma_1,  \bfr_2, \sigma_2 }$~\cite{berry_phase, opes_1,opes_2,Chang_2014}
, where $(\bfr_1, \sigma_1)$  and $(\bfr_2, \sigma_2)$ are restricted in the subsystem $A$. When a 2D translationally invariant system is considered, we can divide our system into two parts along $x$-direction, and keep $y$-direction translationally invariant by applying a periodic boundary condition. The resulting Hamiltonian is then block-diagonal in terms of the wave number $k_y$, i.e. 
$ H = \sum_{k_y} H_{k_y} $, where $H_{k_y}$ is a 1D Hamiltonian for each given $k_y$ subspace. Similar calculation along $y$ direction can be also obtained by applying periodic boundary condition in the $x$ direction first. 

After applying periodic boundary condition in one direction (say $y$) only, we could choose interface along the $x$-direction and the entire Green's function matrix of wave number $k_y$ is given by 
\be
G(k_y)_{ x_1, \sigma_1,x_2, \sigma_2} \! = \!  \! \frac{1}{L_x} \!\!   \sum_{k_x,k_y \in BZ}  \!\! e^{i k_x (x_1\! -\! x_2)} \!  G(k_x,k_y)_{\sigma_1,\sigma_2},
\nonumber\\
\ee
where $L_x$ is the number sites in the $x$-direction and $k_x, k_x$ takes values in the first Brillouin zone.
$G(k_x,k_y)_{\sigma_1,\sigma_2}$  is a $4 \times 4$ matrix determined by the eigenvectors of mean-field Hamiltonian Eq.~\ref{H_original_2}. 
We numerically diagonalize the block's Green's function for the subsystem $A$ with a finite size, e.g., $L_{A}=20$ and $L=40$ as shown in Fig.~\ref{fig:structure} (b).  
The numerical results converges and is independent of the choices of subsystem in the thermal dynamic limit, $L\gg L_A\gg 1$.

\subsection{Topological edge mode}
Besides the calculation of entanglement spectrum of a bulk, we also investigate the edge state properties for the bilayer system numerically. 
According to the bulk-edge correspondence, a nontrivial bulk topological phase should imply the existence of edge states inside the gap. 
In order to study these edges states more clearly in the 2D systems, we first do Fourier transform along a direction parallel to the edge to get a family of 1D Hamiltonians, which are parametrized by the wave number along the edge. 
More precisely, for example, we can make Fourier transform along the $y$-direction first for the field operator,  
$c_{x,y,\sigma}^{} = \frac{1}{\sqrt {L_y}}  \sum_{k_y} e^{iyk_y} c_{x,k_y,\sigma}^{} $ with $L_y$ being the lattice size along the $y$ direction (periodic boundary condition).
An effective BdG Hamiltonian can then be easily derived to be $H_{\text{BdG}}= \sum_{k_y} H_{k_y}$, where
\begin{align}
H_{k_y}  = 
 \sum_{x,\sigma=\uparrow,\downarrow}   
 &
\Big[  \;  (-\mu -  2 t  \cos k_y )  c_{x,k_y,\sigma}^{\dagger} c_{x,k_y,\sigma}^{  }   \\
 -  t c_{x,k_y,\sigma}^{\dagger}& c_{x+1,k_y,\sigma}^{  } 
+e^{i\phi_{\sigma}} \Delta_p  c_{x,k_y,\sigma}^{\dagger}c_{x+1,-k_y,\sigma}^{  \dagger}   \notag \\
 +i e^{i\alpha_{\sigma}}&  e^{i\phi_{\sigma}}  \Delta_p  e^{ik_y}   c_{x,k_y,\sigma}^{\dagger} c_{x,-k_y,\sigma}^{\dagger} +h.c. \Big]      \notag \\
   +\sum_{x} \Big[ -t_z & c^{\dagger}_{x,k_y,\uparrow}c^{}_{x,k_y,\downarrow}
   +     \Delta_s  c^{\dagger}_{x,k_y,\uparrow}c^{\dagger}_{x,-k_y,\downarrow} + h.c. \Big]  \notag 
\end{align} 
One can see that 
\begin{align}
H_{k_y}^\text{4-leg}= H_{k_y} +H_{-k_y}
\end{align} 
is a four-leg ladder system for a given momentum $k_y$. We could then apply a finite size exact diagonalization in the $x$-direction, making the original 2D system mapped to an effective 1D 4-leg ladder system. 
Since the system has two boundary edges at $x=1$ and $x=L_x$, we could then solve the eigenmode spectrum of this finite size system as a function of the momentum $k_y$ numerically.

\section{Results for two examples}

In general, a  phase transition between different phases occur only when the energy gap of the bulk spectrum closes.
Thus, to identify parameter regions for which different (topological phases) are realized, we would first examine the bulk spectrum of the system.
To obtain the bulk spectrum, we would adopt periodic boundary condition (PBC) to form bulk Hamiltonian and study the phase boundary. 
On the other hand, from the bulk-edge correspondence, a nontrivial bulk topological number implies the existence of gapless edge states. 
To investigate edge states, we would consider the open boundary condition along one direction to form a family of 1D Hamiltonians parametrized by wave number $k$. For example, by solving numerically the energy spectrum as a function of the momentum $k_y$ in the y direction, we could study edge states 
near the interface.
By tuning parameters of the Hamiltonian, the phase transition will occur. 
We may classify these phases into  two types according to their energy gap of the bulk spectrum and topological order:
Type I: topological order phase with  symmetry $\mathcal{P}$ by fixed  $t_z\neq 0$ and  $(\phi_{\uparrow},\phi_\downarrow)=(0,\pi), (\alpha_{\uparrow},\alpha_\downarrow) =(0,\pi) $  .   
Type II:  topological order phase with symmetries, $\mathcal{P,T,C} $ 
by fixed  $t_z= 0$ and  $(\phi_{\uparrow},\phi_\downarrow)=(0,0), (\alpha_{\uparrow},\alpha_\downarrow) =(0,\pi) $

\subsection{Type I: $(\phi_{\uparrow},\phi_\downarrow,\alpha_{\uparrow},\alpha_\downarrow)=(0,\pi,0,\pi)$. }
\label{sec:results_typeI}

\begin{figure*}[tbp]
\includegraphics[width=0.95\textwidth]{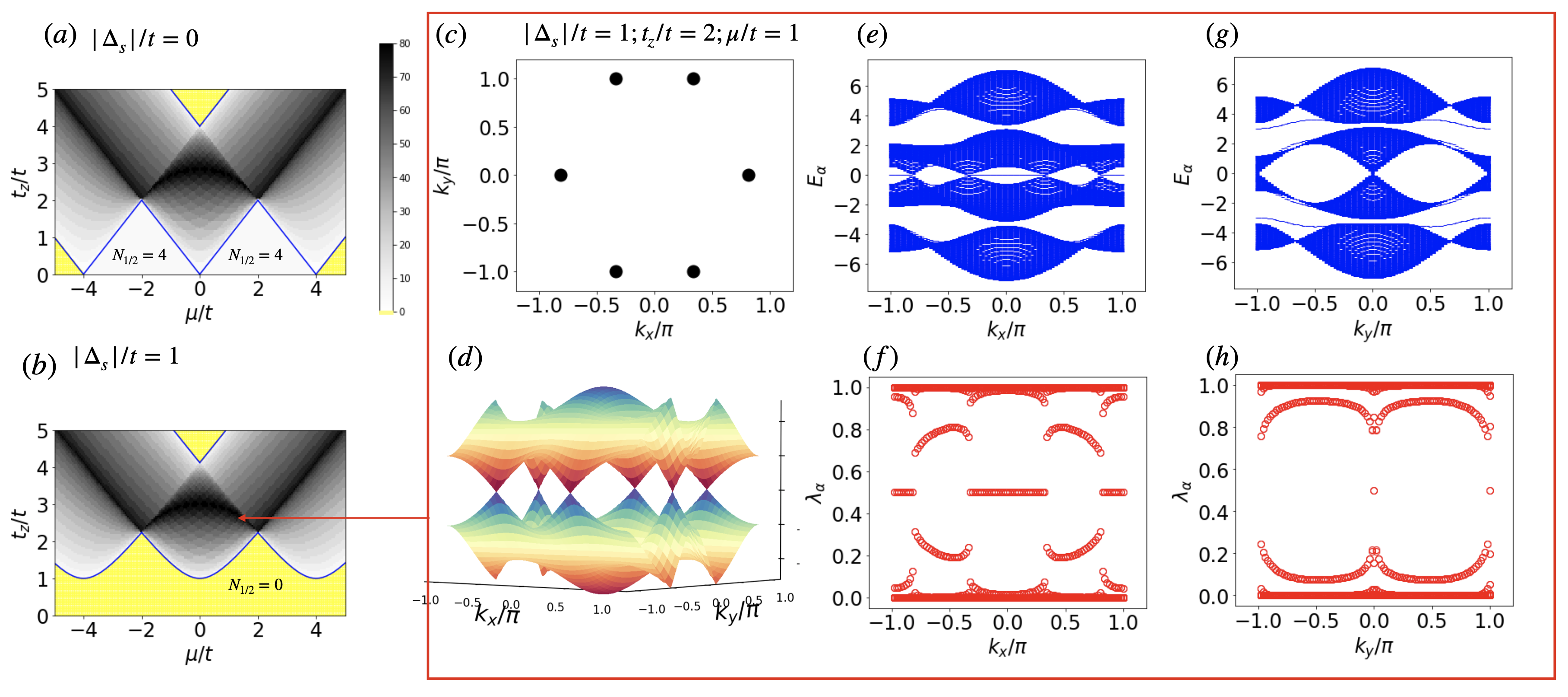}
  \caption{ 
The numerical results $N_{1/2}$ of Hamiltonian Eq.~(\ref{H_original_2}) for $ ( \alpha_{\uparrow}, \alpha_{\downarrow} )=( 0, \pi)$ and  $(\phi_{\uparrow}, \phi_{\downarrow}) = (0,\pi$) with 
(a) $ |\Delta_s|/t=0$,  (b) $ |\Delta_s|/t=1$. 
 $N_{1/2}$ is the number of the degenerate entanglement spectrum at $\lambda_{\alpha}=1/2$ with lattice size $40 \times 40$ and cut system along x-direction. 
 In particular, the result of $N_{1/2}=0$ is marked in yellow. 
 The  blue line shows the boundary between gapped and gapless state, which is determined by energy spectra.
 And, the gapless states with finite $N_{1/2}$ is in grey region. 
  (c) (d)The bulk gap closes at particular momentum $k_x^*$ and $k_y^*$  for $\mu/t=1$, $|\Delta_s|/t=1$, and $t_z/t=2$  with finite size $L=100$.  
  (e) The energy spectra as a function of  $k_x$ (the momentum in x direction), $k_x \in (-\pi,\pi]$ with finite size $L=100$ with boundaries at $y=1$ and $y=100$ for $|\Delta_s|/t=1$, $t_z/t=2$ and  $\mu/t=1$. 
(f) The corresponding spectra of the reduced density with cut at $y=1$ and $y=50$.  
(g)The energy spectra and (h) the entanglement  spectra as a function of $k_y$. }  
\label{fig:typeI_b}
\end{figure*}

\begin{figure}[tbp]
\includegraphics[width=0.46\textwidth]{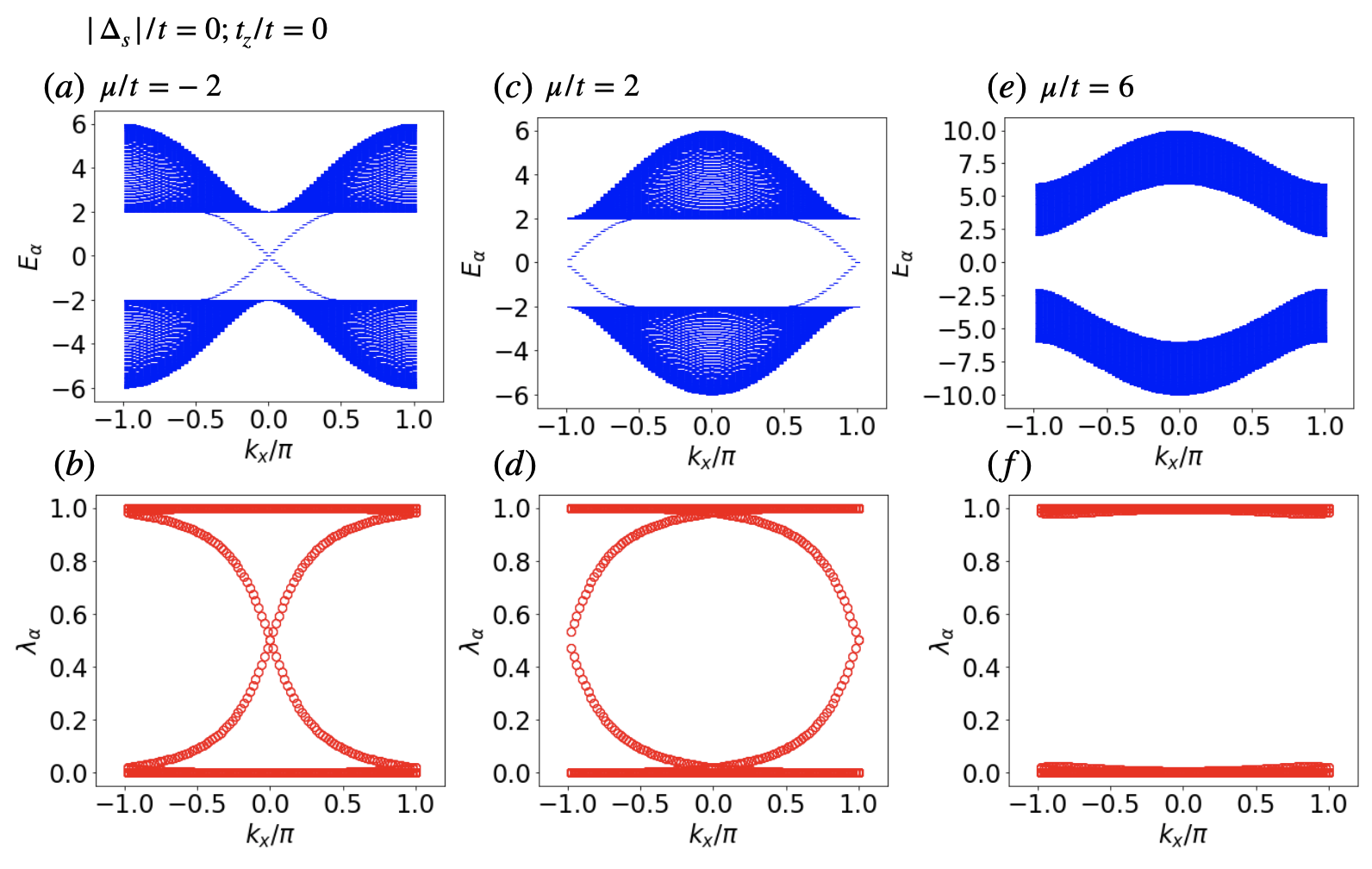}
  \caption{ 
  The energy spectra and the corresponding entanglement spectra as a function $k_x$ with finite size $L = 100$ for 
  $(\phi_{\uparrow},\phi_\downarrow)=(0,\pi), (\alpha_{\uparrow},\alpha_\downarrow) =(0,\pi) $, $t_z/t=0$ and $|\Delta_s|/t=0$ with (a)(b) $\mu/t=-2$,  (c)(d) $\mu/t=2$, and (e)(f) $\mu/t=6$.
}   
\label{fig:typeI_a}
\end{figure}

As a check and benchmark, we first calculate the the energy eigenvalues of the bulk Hamiltonian for such a anti-symmetric case with 
 $(\phi_{\uparrow},\phi_\downarrow)=(0,\pi)$ and 
$(\alpha_{\uparrow},\alpha_\downarrow) =(0,\pi) $.
As explained above, depending on the  parameter $t_z/t$ and $|\Delta_s|/t$, the corresponding phase will make a transition to either topological order phase or trivial phase. 
In the following, we would show the details for energy spectra and entanglement spectra.
 A simple evaluation shows that the eigenvalues of matrix  in Eq.~(\ref{H_original_2}) with $(\phi_{\uparrow},\phi_\downarrow)=(0,\pi)$ and 
$(\alpha_{\uparrow},\alpha_\downarrow) =(0,\pi)$ 
are simply $\pm E_{\bfk}$, where 
 the energy dispersion for this system are
\be
 E_{\bfk} =  \pm \Big[ \varepsilon_{\bfk }^2  + t_z^2  + 4  |\Delta_p| ^2  \left( \sin^2 k_x   + \sin^2 k_y \right) 
  +|\Delta_s|^2 \nonumber \\ 
  \pm 2  |t_z|  \sqrt{ \varepsilon_{\bfk }^2  +|\Delta_s|^2 +  4 |\Delta_p|^2  \sin^2 k_x }   \;   \Big]^{1/2} \nonumber.
 \ee
We discuss two cases: $|\Delta_s|/t=0$ ( see Fig.~\ref{fig:typeI_b}(a) ) and  $|\Delta_s|/t=1$ ( see Fig.~\ref{fig:typeI_b}(b)).
The phase diagram can be obtained from computing the number $N_{1/2}$ of the $\lambda_\alpha=1/2$ modes in entanglement spectrum from
the Green's function matrix.
In our case, we numerically diagonalize the Green's function for subsystem $A$ with a finite size, e.g. $L_y=20$ of  full lattice size $40 \times 40$. 
When there is an extensive number of the $\lambda_\alpha=1/2$ mode, it indicates an extensive number of the entangling boundary
modes localized at the interface between subsystem $A$ and subsystem $B$. These entanglement boundary modes mimic the dispersionless 
boundary modes of the gapless topological phases, such as the flat bands on the zigzag edge of graphene \cite{Chang_2014}.  
As shown in Figs. ~\ref{fig:typeI_b} (a)(b), the "large" number $N_{1/2}$ indicates the gapless region and the phase boundaries
between the gapped and the gapless regions can be shown from the intensity plot.

In this model, the gap of the system for $|\Delta_s|/t\!=\!0$ closes only when the following condition is satisfied:
\begin{align}
\label{eq:gap}
4 |\Delta_p|^2 \sin^2 k_y = \Big(  \sqrt{ \varepsilon_{\bfk }^2+4 |\Delta_p|^2  \sin^2 k_x } \pm t_z   \Big)^2=0.
\end{align}
Therefore, the gap closes at $k_y = 0$ or $k_y=\pi$. 
From a straightforward calculation, it is found that this condition is equivalent to
$(-\mu/t \mp 2)^2 +4-2(-\mu/t \mp2) \cos k_x = (t_z/t)^2$, where ``$-$'' term corresponds to $k_y=0$, ``$+$'' term corresponds to $k_y=\pi$, and $k_x$ is related to the lattice size. 
This is, the gap of the system closes at $\bfk$ which is determined by Hamiltonian parameters.

First, we consider the parameters with $|\Delta_s|/t=0$ and $t_z/t=0$, then the Hamiltonian reduces to a bilayer chiral p-wave superconductor without interlayer interaction. 
It can be reduced to 2d chiral p-wave superconductor model. 
There are  three phases separated by three quantum critical points at $\mu/t=0, \pm4$, which are labeled by the Chern number (Ch) as  Ch$=0 \; (|\mu/t|>4)$,   Ch$=-1 \; (-4<\mu/t < 0)$, and  Ch$=1 \; (0< \mu/t <4)$.
 Since in the torus geometry, there are two entangling boundaries that will lead to an additional twofold degeneracy together with
the twofold degeneracy from the bilayer of the system
in the entanglement spectrum.  The number of the $\lambda_\alpha=1/2$ in the entanglement spectrum is four times the Chern number,
$N_{1/2}= 4 |{\rm Ch}|$. I,e, $N_{1/2}=4$ when the Chern number is unity (Ch$=\pm1$).
Our numerical results agree with  the analytic result of the Chern number, confirming the topological properties in our current system.
There are two phases which are labeled by by $N_{1/2}$ as $N_{1/2}=4$ (topological phase) and $N_{1/2}=0$ (trivial phase).

We shall now look more carefully into the edge states. 
We impose the periodic boundary condition in the $x$ direction. 
Fourier transforming along the $x$ direction and obtain a family of 1D Hamiltonians parametrized by $k_x$.
Suppose that the system has two edges at $y=1$ and $y=L$. 
By solving numerically the energy spectrum as a function of the momentum $k_x$ in the y direction, we then study edge states.
Figs.~\ref{fig:typeI_a}(a)(c)(e) show the energy spectra with open boundaries along $y$ axis and transnational insurance along the $x$ direction. 
It shows zero energy modes appear for $k_x=0$ or $k_x= \pi$.
On the other hand,  the entanglement spectrum can mimic the physical edge spectrum
for the torus geometry. The corresponding edge modes in the entanglement  spectrum will be localized at the entangling boundary
between the subsystem $A$ and the subsystem $B$, and the entanglement spectra are shown in  Figs.~\ref{fig:typeI_a} (b)(d)(f).
These results are consistent with the bulk- edge correspondence again.
The Figs~\ref{fig:typeI_a}(a)(c) show that these points are topological phase with $N_{1/2}=4$. 
In $\mu/t=2$, the fourfold degeneracy occur at $k_x=0$.  
In $\mu/t=-2$, the fourfold degeneracy occur at $k_x=\pi$.  
The $N_{1/2}$'s appear at different $k_x$ corresponding to the different topological phases.

\begin{figure*}[htpb]
\includegraphics[width=0.95\textwidth]{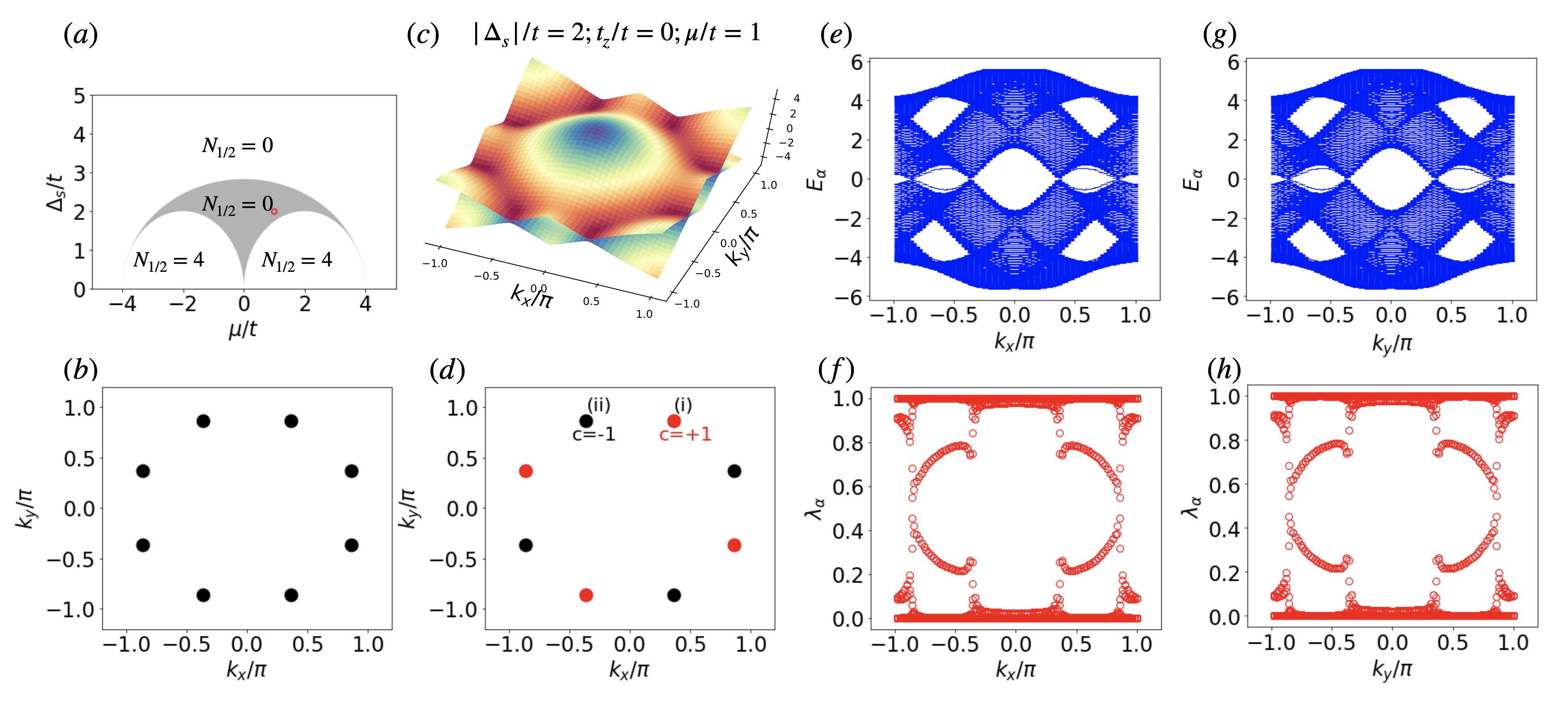}
  \caption{
  (a)  The phase diagram  as computed via the  energy eigenvalue for $t_z/t=0$, $\phi_{\uparrow}=\phi_{\downarrow} =0$ and  $ \alpha_{\uparrow}=0  \;  \alpha_{\downarrow}= \pi$ with under periodic boundary condition. 
Shaded regimes are gapless phases.
For $\mu/t=1$, $|\Delta_s|/t=2$, and $t_z/t=0$, 
 (b) The bulk gap closes at particular momentum $k_x^*$ and $k_y^*$ (belong to $C_{4v}$ group) with finite size $L=100$. 
 (c) The lower energy spectra  and (d) topological charge of the function of momentum $k_x$ and $k_y$ 
 The energy spectra as a function  (e) $k_x$ and (g) $k_y$ with finite size L = 100 and the corresponding entanglement spectra as a function of (f) $k_x$ and (h) $k_y$. }   
\label{fig:typeII}
\end{figure*}

 Next, we consider the case $|\Delta_s|/t \neq 1$. 
In Fig.~\ref{fig:typeI_b} (b), we find there exist a gapless phase in the phase diagram.
The gap closing points $(k_x^* , k_y^*)$  are determined by Eq. (\ref{eq:gap}) with $|\Delta_s|/t=1$, $|t_z|/t=2$, $|\mu|/t=1$ in Figs.~\ref{fig:typeI_b}(c)(d).
We then impose the periodic boundary condition in the both $x$ direction and  $y$ direction. 
The energy spectra with open boundary condition as a function of momentum are shown in Fig.~\ref{fig:typeI_b}(e)(g). 
The corresponding spectra of the reduced density with open ends also is shown in Fig.~\ref{fig:typeI_b}(f)(h).
It shows zero energy modes appear between the four gapless points $k_x^*$. 
These results are consistent with the bulk-edge correspondence for the topological gapless phases \cite{Matsuura_2013}.



\subsection{Type II: $(\phi_{\uparrow},\phi_\downarrow,\alpha_{\uparrow},\alpha_\downarrow)=(0,0,0,\pi)$. }
\label{sec:results_typeII}

We consider the fixed parameters with $(\phi_{\uparrow},\phi_\downarrow)=(0,0), (\alpha_{\uparrow},\alpha_\downarrow) =(0,\pi) $   at $ t_z/t =0$. 
In this case, the mean-feild Hamiltonian has parity and time reversal symmetry. 
In particular, this model can be transformed to Kondo lattice model in some special case~\cite{Kondo_Lattice}, which has the time-reversal symmetry and rotation symmetry.
In Ref~\cite{Kondo_Lattice}, they studied  a topological transition from a strong topological insulator to a weak topological insulator in a three-dimensional Kondo lattice. 
They showed the transitions between two topological phases must go through gapless phases. 
In this section, we would show the similar phase transition in the bilayer system. 

First, it is easy to check that for such a anti-symmetric case with fixed $t_z/t=0$, 
the energy dispersion for this system are
\be 
E_{\bfk}= \pm \! \left[    \varepsilon_{\bfk}^2 +    \Big( |\Delta_s| \pm 2 |\Delta_p|   \sqrt{ \sin^2 k_x + \sin^2 k_y } \Big)^2     \right] ^{1/2} \nonumber \ee
as expected. Again, the energy gap of the system closes only when the following condition is satisfied:
\be  \varepsilon_{\bfk}^2 = \Big( |\Delta_s| \pm 2 |\Delta_p|  \sqrt{ \sin^2 k_x + \sin^2 k_y } \Big)^2 = 0. \nonumber \ee
It is found that this condition is equivalent to
\be
&& \cos^2 {k_x} + \cos^2 {k_y} = 2- |\Delta_s|^2/4t^2,  \nonumber \\
&& \cos {k_x} + \cos {k_y} = -\mu/2t.
\label{gapclose_tz0_c_1}
\ee
The gap closing points are shown in Fig.~\ref{fig:typeII}(a). 
It is obvious to see that if $\bfk=(k_x,k_y)$ is a generic solution, $(\pm k_x,\pm k_y)$ and $(\pm k_y,\pm k_x)$ are also solutions, i.e. there are eight-fold degeneracy, belong to $C_{4v}$ group (see Fig.~\ref{fig:typeII}(b)).
In Fig.~\ref{fig:typeII}(c), we show the case with  $\mu/t=1$, $|\Delta_s|/t=2$, and $t_z/t=0$ numerically. 
On the other hand,  we also calculate the entanglement spectrum and we find that there are regions  with a large number of $N_{1/2}$
in the entanglement spectrum. These regions are the gapless phases which is same as the previous analysis. 

However, in this case, we do not find the Majorana zero modes at each end in gapless region (see Figs.~\ref{fig:typeII}(e)(g)) for the open boundary conditions
in both $x$ and $y$ directions. 
For the entanglement spectrum, we also do not find any entangling boundary modes with $\lambda_\alpha=1/2$. 
The absence of Majorana edge modes does not indicate it is a trivial gapless phase. 
Duo to the perfect cancellation of the topological charges of the gapless points in the bulk projected on the boundary, the Majorana zero modes are gapped out.

Nevertheless, 
we can directly calculate the  topological charge which is  given by the winding number that corresponds to eigenstate of effective Hamiltonian of a gapless point.  
To compute the effective Hamiltonian, We first expand the Hamiltonian
around the gapless points $(k_{x}^{0},k_{y}^{0})$ up to linear terms, i.e.
$ \bfk= \bfk^{0}+ \bfq $, so that  
$H ( \bfk ) \sim H _{0}(\bfk^{0})+ H _{1}(\bfq )$, where
$$
\begin{aligned}
&H_{0}\left( \bfk^{0}\right)
=\varepsilon_{\bfk} \left(\sigma_{3} \otimes \sigma_{0}\right)-\Delta_{s}\left(\sigma_{2} \otimes \sigma_{2}\right)\\
&\quad-2 \Delta_{p}\left[\sin \left(k_{x}^{0}\right)\right]\left(\sigma_{2} \otimes \sigma_{0}\right)-2 \Delta_{p}\left[\sin \left(k_{y}^{0}\right)\right]\left(\sigma_{1} \otimes \sigma_{3}\right)
\end{aligned}
$$
and
$$
\begin{aligned}
&H_{1}(\bfq)=2 t [ {q_{x}} \sin \left(k_{x}^{0}\right)+{q_{y}}  \sin \left(k_{y}^{0}\right)]\left(\sigma_{3} \otimes \sigma_{0}\right)\\
&-2 \Delta_{p}{q_{x}} \cos \left(k_{x}^{0}\right)\left(\sigma_{2} \otimes \sigma_{0}\right)-2 \Delta_{p}{q_{y}} \cos \left(k_{y}^{0}\right)\left(\sigma_{1} \otimes \sigma_{3}\right).
\end{aligned}
$$

Since the analytic solutions of $H_0$ are not generally available, we first numerically calculate its two eigenvectors, $|v_{1}\rangle$ and  $|v_{2}\rangle$, with zero eigenenergy at  $E_{k}=\pm 0$. Here, we just consider gapless points (see Fig.~\ref{fig:typeII}(b) ) at  $ (\alpha_{\uparrow},\alpha_{\downarrow} ) =(0,\pi)$,  
    $ (\phi_{\uparrow},\phi_{\downarrow}) =(0,0)$, $\mu / t=1$,$\left|\Delta_{s}\right| / t=2$, 
and $t_{z} / t=0$. 
The effective Hamiltonian near the gapless points is then obtained by
projecting the leading order Hamiltonian, $H_{1}(\bfq)$, onto the subpsace of $\{|v_{1}\rangle,|v_{2}\rangle\}$, i.e.
\begin{eqnarray}
H _{\rm eff }( \bfq )&=\left(\begin{array}{cc}
\left\langle v_{1}\left|H_{1}( \bfq)\right| v_{1}\right\rangle & \left\langle v_{1}\left|H_{1}(\bfq)\right| v_{2}\right\rangle \\
\left\langle v_{2}\left|H_{1}(\bfq)\right| v_{1}\right\rangle & \left\langle v_{2}\left|H_{1}(\bfq)\right| v_{2}\right\rangle
\end{array}\right) \notag \\
&= v_x({\bf k}^0) q_x \sigma_x' + v_y({\bf k}^0) q_y \sigma_y',
\label{Eq:H_eff new basis}
\end{eqnarray}
Here $v_x({\bf k}^0)$ and $v_y({\bf k}^0)$ are two "velocity functions" of the gapless point. Note that $\sigma_{x/y}'$ are Pauli matrix expressed in another basis, which are rotated from $\{|v_{1}\rangle,|v_{2}\rangle\}$, in order to simplify the notation to the conventional Wyel point form in the 2D system.

Note that, within this new basis, the chiral symmetry is given by $\sigma_z'$, i.e. $\sigma_z' H _{\rm eff }( \bfq )  \sigma_z' = - H _{\rm eff }( \bfq )$.
As a result, we can further rescale the effective Hamiltonian and express it into polar coordinate:
\begin{eqnarray}
H_{\rm eff }( \bfq ) &=& E_{1}(\bfq)
\left(\begin{array}{cc}
0 & e^{i \theta_{\bfq}} \\e^{-i \theta_{\bfq}} & 0
\end{array}\right),
\end{eqnarray}
where $E_1(\bfq) \equiv \sqrt{v_x(\bfk^0)^2{q}_x^2+v_y(\bfk^0)^2{q}_y^2}$ and $\theta_\bfq \equiv \tan^{-1} \left[\frac{v_y(\bfk^0){q}_y}{v_x(\bfk^0){q}_x}\right]$.
The corresponding winding number could then be calculated to be
\begin{align}
W = \frac{1}{2 \pi i} \oint d\theta_{\tilde{\bf{q}}} \partial_{\theta_{\tilde{\bf{q}}}} e ^{i \theta_{\tilde{\bf{q}}}} =1.
\end{align} 
The topological charge of the gapless points at point (ii) (see  Fig.~\ref{fig:typeII}(d) )    is $W=-1$, since one of velocity function ($v_x(\bfk^0)$ and $v_y(\bfk^0)$) changes its sign in Eq. (\ref{Eq:H_eff new basis}). 
Following similar approach, we can get all gapless points's topological charge shown in the Fig.~\ref{fig:typeII}(d).

\section{Evidence in Time-of-flight experiment}
\label{sec:discussion}

In the literature, there have been several proposals to measure Majorana modes. In condensed matter systems, the chiral Majorana fermions in two dimensions can be measured from the quantized thermal Hall conductivity~\cite{Read_2000, Sumiyoshi_2013, Nomura_2012} or from the scanning tunneling microscopy \cite{Chang_2014_2, Yin_2015, Palacio-Moraleseaav6600, Frolov_2020, Wang333}. In cold-atom experiments, an alternative detection of the Majorana modes is from the long-distance Majorana correlation from the time-of-flight (ToF) experiments \cite{Zoller_1Dladder}.

However, in the gapless topological superfluid proposed here, the Majorana zero modes appear when the relative gauge phases between the inter-layer $s-$wave pairing and the intra-layer $p-$wave pairing order parameters are changed from their conventional value, $(\alpha_\uparrow,\alpha_\downarrow,\phi_\uparrow,\phi_\downarrow)=(0,0,0,0)$. According to Table \ref{table:symmetry}, the most generic Majorana modes should appear at Type I, where  $t_z\neq 0$ and $(\alpha_\uparrow,\alpha_\downarrow,\phi_\uparrow,\phi_\downarrow)=(0,\pi,0,\pi)$.

Since such kind of many-bidy gauge phase may not be easily manipulated, it becomes important to how to measure such spontaneously emergent gauge phase in the present experiment. According to the calculated positions of the associated gapless points, see Fig. \ref{fig:typeI_b} Type I case, the low energy spectrum are highly asymmetric in the $k_x$ and $k_y$ direction due to the "six-fold" gapless point structure shown in (c). It is therefore interesting if one could observe this in the simple ToF experiment too.

It is known that the time-of-flight image for a free expansion time $t$ is obtained by integrating over the $z$ direction, i.e.,
\begin{eqnarray}
n_{ToF}(x,y) &\propto & 
\int_{-\infty}^{\infty} dz \; n_{m {\bf r}/\hbar t }
\left|\tilde{w}_{0}\left(\frac{m{\bf r}}{\hbar t}\right)\right|^2
\label{Eq:n_ToF}
\end{eqnarray} 
where $\tilde{w}_{0}(m{\bf r}/\hbar t)$ is the lowest band Wannier function of the optical lattice in momentum space. The final position ${\bf r} = \hbar {\bf k} t /m$ is to be transformed into the initial momentum under the assumption of long time-of-flight and free expansion. $n_{\bf k}$ is the momentum distribution {\it inside } before expansion,
\begin{eqnarray}
n_{\bf k} &=& \sum_{{\bf j},{\bf j}'} \sum_{\sigma,\sigma'}  \langle G | c_{{\bf j},\sigma}^{\dagger}   c_{{\bf j}',\sigma}  | G \rangle e^{i(j_x-j_x')k_xa }e^{i(j_y-j_y')k_ya }
\nonumber\\
&&\times e^{i(\sigma-\sigma')k_zd}
\end{eqnarray} 
where $ | G \rangle$ is the ground-state wave function obtained by diagonalizing the mean-field Hamiltonian. ${\bf j}=(j_x,j_y)$ is the lattice index in the 2D plane and $\sigma=\pm 1/2$ for layer index.

\begin{figure}[htpb]
\includegraphics[width=0.45\textwidth]{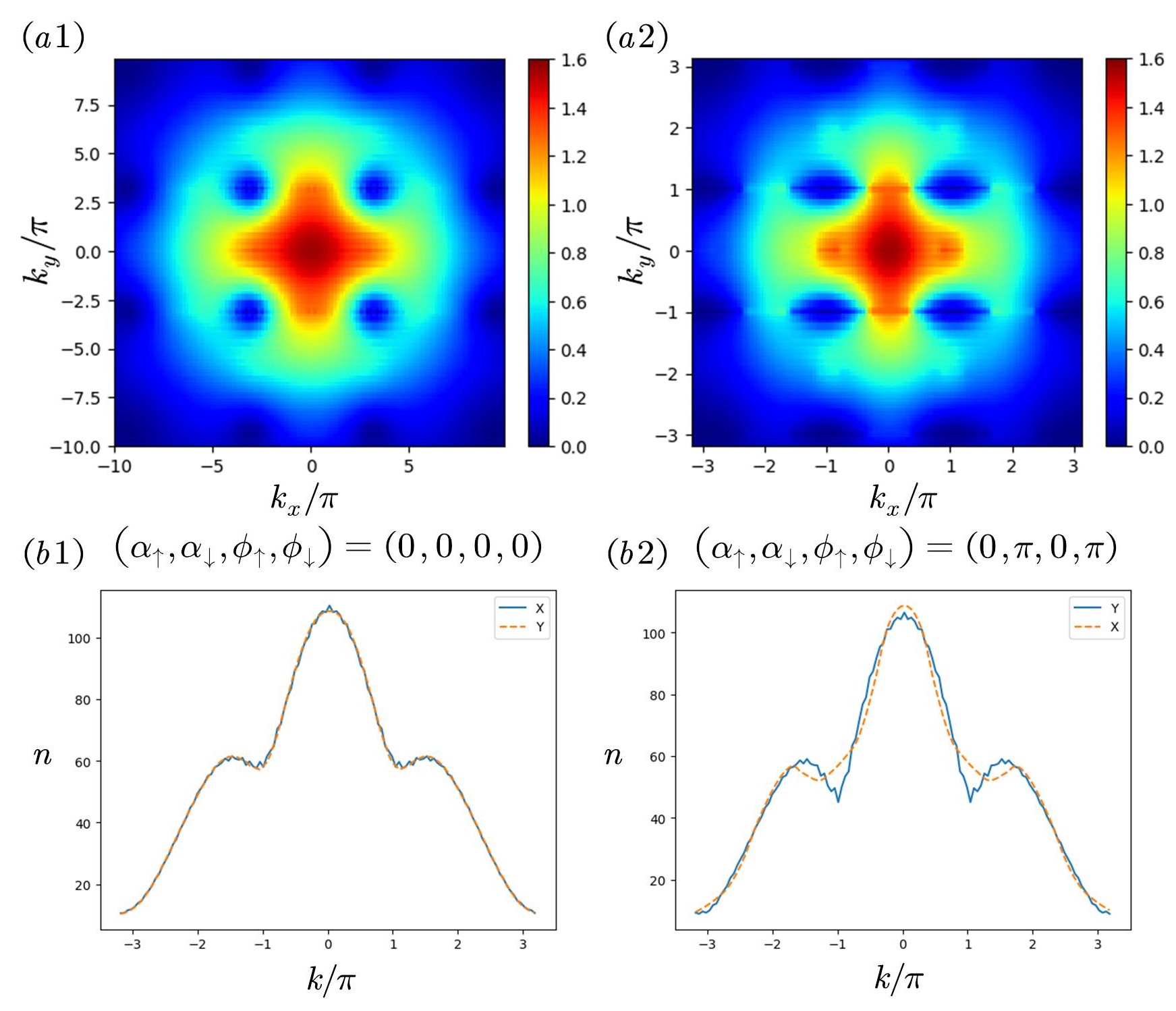}
  \caption{
(a1) and (a2) are the calculated ToF images for two different gauge phases,  $(\alpha_\uparrow,\alpha_\downarrow,\phi_\uparrow,\phi_\downarrow)=(0,0,0,0)$ and $(\alpha_\uparrow,\alpha_\downarrow,\phi_\uparrow,\phi_\downarrow)=(0,\pi,0,\pi)$. (b1) and (b2) are the integrated density distribution, $n_{ToF}(x,y)$, for (a1) and (a2) respectively. Both direction in $x$ and $y$ direction are shown together for comparison.
}   
\label{fig:TOF}
\end{figure}

In Fig.~\ref{fig:TOF}, we show the TOF images for the system with $t_z/t=2$, $\mu/t=0$, and $|\Delta_s|/t=1$ in a bilayer system of $L_x=L_y=100$ (open boundary condition). Results obtained by both
$(\alpha_\uparrow,\alpha_\downarrow,\phi_\uparrow,\phi_\downarrow)=(0,0,0,0)$ and  $(\alpha_\uparrow,\alpha_\downarrow,\phi_\uparrow,\phi_\downarrow)=(0,\pi,0,\pi)$ are shown togther in (a1) and (a2) respectively. These ToF images are taken by integrating along the $z$ axis (perpendicular to the layer) after long time of flight. One could see that is a deep valley at certain finite value in the $y$ axis.

In order to investigate results of these two different gauge phases more specifically, in (b1) and (b2) we further show their integrated density distribution both in the $x$ and $y$ directions for comparison. We could see that in the regime of gapless topological phase (b2) (same as the case in Fig. \ref{fig:typeI_b}), the ToF images show a significant difference in the $x$ and $y$ directions, while no such difference in the conventional phases shown in (b1). The most significant difference in their distribution appears at $k_ya=\pm \pi$, or equivalently at $y=\pm \pi\hbar t/ma$. This is obvious due to the position of double gapless Weyl points at $k_y=\pm \pi/a$ as shown in Fig. \ref{fig:typeI_b}(c).
It is because at 2D system, the presence of gapless points makes the density of states almost zero in the low energy regime, while no such a similar situation in the $k_x$ direction. On the other hand, when considering the system with the conventional gauge phase in (a1) and (b1), the system is a gaped phase and therefore not much difference in the $x$ and $y$ directions. Therefore, our results show that the phase modulation of the gapless topological superfluid could also be observed easily by analyzing the ToF images.

\section{Conclusion}
\label{sec:conclusion}

In this paper, we propose that a topological state can be prepared and observed in a paired 2D $p$-wave superfluids, where spin polarized fermionic atoms are loaded in a bilayer optical lattice with a weak coupling to Rydberg excited states
The associated quantum degenerate Majorana zeros modes appear in the edge, confirmed by the zero energy mode and entanglement spectrum numerics. 
Our work paves the way for future investigation of the many- body physics with non-Abelian statistics under time-reversal symmetry.

\acknowledgements
We thank useful discussion with Chung-Hou Chung, Yi-Ping Huang, Chung-Yu Mou, and Jhih-Shih You.
CYH is  supported by the MOST of Taiwan under Grants No. 108-2112-M-029 -006 -MY3. 
DWW is supported by National Center for theoretical Sciences and by the Higher Education Sprout Project funded by the Ministry of Science and Technology and Ministry of Education in Taiwan.


\begin{thebibliography}{68}%
\makeatletter
\providecommand \@ifxundefined [1]{%
 \@ifx{#1\undefined}
}%
\providecommand \@ifnum [1]{%
 \ifnum #1\expandafter \@firstoftwo
 \else \expandafter \@secondoftwo
 \fi
}%
\providecommand \@ifx [1]{%
 \ifx #1\expandafter \@firstoftwo
 \else \expandafter \@secondoftwo
 \fi
}%
\providecommand \natexlab [1]{#1}%
\providecommand \enquote  [1]{``#1''}%
\providecommand \bibnamefont  [1]{#1}%
\providecommand \bibfnamefont [1]{#1}%
\providecommand \citenamefont [1]{#1}%
\providecommand \href@noop [0]{\@secondoftwo}%
\providecommand \href [0]{\begingroup \@sanitize@url \@href}%
\providecommand \@href[1]{\@@startlink{#1}\@@href}%
\providecommand \@@href[1]{\endgroup#1\@@endlink}%
\providecommand \@sanitize@url [0]{\catcode `\\12\catcode `\$12\catcode
  `\&12\catcode `\#12\catcode `\^12\catcode `\_12\catcode `\%12\relax}%
\providecommand \@@startlink[1]{}%
\providecommand \@@endlink[0]{}%
\providecommand \url  [0]{\begingroup\@sanitize@url \@url }%
\providecommand \@url [1]{\endgroup\@href {#1}{\urlprefix }}%
\providecommand \urlprefix  [0]{URL }%
\providecommand \Eprint [0]{\href }%
\providecommand \doibase [0]{https://doi.org/}%
\providecommand \selectlanguage [0]{\@gobble}%
\providecommand \bibinfo  [0]{\@secondoftwo}%
\providecommand \bibfield  [0]{\@secondoftwo}%
\providecommand \translation [1]{[#1]}%
\providecommand \BibitemOpen [0]{}%
\providecommand \bibitemStop [0]{}%
\providecommand \bibitemNoStop [0]{.\EOS\space}%
\providecommand \EOS [0]{\spacefactor3000\relax}%
\providecommand \BibitemShut  [1]{\csname bibitem#1\endcsname}%
\let\auto@bib@innerbib\@empty
\bibitem [{\citenamefont {Chou}\ \emph {et~al.}(2016)\citenamefont {Chou},
  \citenamefont {Zhai}, \citenamefont {Chung}, \citenamefont {Mou},\ and\
  \citenamefont {Lee}}]{Kondo_Lattice}%
  \BibitemOpen
  \bibfield  {author} {\bibinfo {author} {\bibfnamefont {P.-H.}\ \bibnamefont
  {Chou}}, \bibinfo {author} {\bibfnamefont {L.-J.}\ \bibnamefont {Zhai}},
  \bibinfo {author} {\bibfnamefont {C.-H.}\ \bibnamefont {Chung}}, \bibinfo
  {author} {\bibfnamefont {C.-Y.}\ \bibnamefont {Mou}},\ and\ \bibinfo {author}
  {\bibfnamefont {T.-K.}\ \bibnamefont {Lee}},\ }\bibfield  {title} {\bibinfo
  {title} {Emergence of a fermionic finite-temperature critical point in a
  kondo lattice},\ }\href {https://doi.org/10.1103/PhysRevLett.116.177002}
  {\bibfield  {journal} {\bibinfo  {journal} {Phys. Rev. Lett.}\ }\textbf
  {\bibinfo {volume} {116}},\ \bibinfo {pages} {177002} (\bibinfo {year}
  {2016})}\BibitemShut {NoStop}%
\bibitem [{\citenamefont {Majorana}(2008)}]{MF}%
  \BibitemOpen
  \bibfield  {author} {\bibinfo {author} {\bibfnamefont {E.}~\bibnamefont
  {Majorana}},\ }\bibfield  {title} {\bibinfo {title} {Teoria simmetrica
  dell'elettrone e del positrone},\ }\href {https://doi.org/10.1007/BF02961314}
  {\bibfield  {journal} {\bibinfo  {journal} {Il Nuovo Cimento (1924-1942)}\
  }\textbf {\bibinfo {volume} {14}},\ \bibinfo {pages} {171} (\bibinfo {year}
  {2008})}\BibitemShut {NoStop}%
\bibitem [{\citenamefont {Chiu}\ \emph {et~al.}(2016)\citenamefont {Chiu},
  \citenamefont {Teo}, \citenamefont {Schnyder},\ and\ \citenamefont
  {Ryu}}]{classification}%
  \BibitemOpen
  \bibfield  {author} {\bibinfo {author} {\bibfnamefont {C.-K.}\ \bibnamefont
  {Chiu}}, \bibinfo {author} {\bibfnamefont {J.~C.~Y.}\ \bibnamefont {Teo}},
  \bibinfo {author} {\bibfnamefont {A.~P.}\ \bibnamefont {Schnyder}},\ and\
  \bibinfo {author} {\bibfnamefont {S.}~\bibnamefont {Ryu}},\ }\bibfield
  {title} {\bibinfo {title} {Classification of topological quantum matter with
  symmetries},\ }\href {https://doi.org/10.1103/RevModPhys.88.035005}
  {\bibfield  {journal} {\bibinfo  {journal} {Rev. Mod. Phys.}\ }\textbf
  {\bibinfo {volume} {88}},\ \bibinfo {pages} {035005} (\bibinfo {year}
  {2016})}\BibitemShut {NoStop}%
\bibitem [{\citenamefont {Cheng}\ \emph {et~al.}(2016)\citenamefont {Cheng},
  \citenamefont {Zaletel}, \citenamefont {Barkeshli}, \citenamefont
  {Vishwanath},\ and\ \citenamefont {Bonderson}}]{Bosonic_SPT}%
  \BibitemOpen
  \bibfield  {author} {\bibinfo {author} {\bibfnamefont {M.}~\bibnamefont
  {Cheng}}, \bibinfo {author} {\bibfnamefont {M.}~\bibnamefont {Zaletel}},
  \bibinfo {author} {\bibfnamefont {M.}~\bibnamefont {Barkeshli}}, \bibinfo
  {author} {\bibfnamefont {A.}~\bibnamefont {Vishwanath}},\ and\ \bibinfo
  {author} {\bibfnamefont {P.}~\bibnamefont {Bonderson}},\ }\bibfield  {title}
  {\bibinfo {title} {Translational symmetry and microscopic constraints on
  symmetry-enriched topological phases: A view from the surface},\ }\href
  {https://doi.org/10.1103/PhysRevX.6.041068} {\bibfield  {journal} {\bibinfo
  {journal} {Phys. Rev. X}\ }\textbf {\bibinfo {volume} {6}},\ \bibinfo {pages}
  {041068} (\bibinfo {year} {2016})}\BibitemShut {NoStop}%
\bibitem [{\citenamefont {Kitaev}(2001)}]{Kitaev_1D}%
  \BibitemOpen
  \bibfield  {author} {\bibinfo {author} {\bibfnamefont {A.~Y.}\ \bibnamefont
  {Kitaev}},\ }\bibfield  {title} {\bibinfo {title} {Unpaired majorana fermions
  in quantum wires},\ }\href {http://stacks.iop.org/1063-7869/44/i=10S/a=S29}
  {\bibfield  {journal} {\bibinfo  {journal} {Physics-Uspekhi}\ }\textbf
  {\bibinfo {volume} {44}},\ \bibinfo {pages} {131} (\bibinfo {year}
  {2001})}\BibitemShut {NoStop}%
\bibitem [{\citenamefont {Fu}\ and\ \citenamefont {Kane}(2008)}]{Proximity_MF}%
  \BibitemOpen
  \bibfield  {author} {\bibinfo {author} {\bibfnamefont {L.}~\bibnamefont
  {Fu}}\ and\ \bibinfo {author} {\bibfnamefont {C.~L.}\ \bibnamefont {Kane}},\
  }\bibfield  {title} {\bibinfo {title} {Superconducting proximity effect and
  majorana fermions at the surface of a topological insulator},\ }\href
  {https://doi.org/10.1103/PhysRevLett.100.096407} {\bibfield  {journal}
  {\bibinfo  {journal} {Phys. Rev. Lett.}\ }\textbf {\bibinfo {volume} {100}},\
  \bibinfo {pages} {096407} (\bibinfo {year} {2008})}\BibitemShut {NoStop}%
\bibitem [{\citenamefont {Sau}\ \emph {et~al.}(2010)\citenamefont {Sau},
  \citenamefont {Lutchyn}, \citenamefont {Tewari},\ and\ \citenamefont
  {Das~Sarma}}]{S_Heterostructures}%
  \BibitemOpen
  \bibfield  {author} {\bibinfo {author} {\bibfnamefont {J.~D.}\ \bibnamefont
  {Sau}}, \bibinfo {author} {\bibfnamefont {R.~M.}\ \bibnamefont {Lutchyn}},
  \bibinfo {author} {\bibfnamefont {S.}~\bibnamefont {Tewari}},\ and\ \bibinfo
  {author} {\bibfnamefont {S.}~\bibnamefont {Das~Sarma}},\ }\bibfield  {title}
  {\bibinfo {title} {Generic new platform for topological quantum computation
  using semiconductor heterostructures},\ }\href
  {https://doi.org/10.1103/PhysRevLett.104.040502} {\bibfield  {journal}
  {\bibinfo  {journal} {Phys. Rev. Lett.}\ }\textbf {\bibinfo {volume} {104}},\
  \bibinfo {pages} {040502} (\bibinfo {year} {2010})}\BibitemShut {NoStop}%
\bibitem [{\citenamefont {Alicea}(2010)}]{tunable_S_device}%
  \BibitemOpen
  \bibfield  {author} {\bibinfo {author} {\bibfnamefont {J.}~\bibnamefont
  {Alicea}},\ }\bibfield  {title} {\bibinfo {title} {Majorana fermions in a
  tunable semiconductor device},\ }\href
  {https://doi.org/10.1103/PhysRevB.81.125318} {\bibfield  {journal} {\bibinfo
  {journal} {Phys. Rev. B}\ }\textbf {\bibinfo {volume} {81}},\ \bibinfo
  {pages} {125318} (\bibinfo {year} {2010})}\BibitemShut {NoStop}%
\bibitem [{\citenamefont {Sato}\ \emph {et~al.}(2010)\citenamefont {Sato},
  \citenamefont {Takahashi},\ and\ \citenamefont {Fujimoto}}]{Spin_Singlet_SC}%
  \BibitemOpen
  \bibfield  {author} {\bibinfo {author} {\bibfnamefont {M.}~\bibnamefont
  {Sato}}, \bibinfo {author} {\bibfnamefont {Y.}~\bibnamefont {Takahashi}},\
  and\ \bibinfo {author} {\bibfnamefont {S.}~\bibnamefont {Fujimoto}},\
  }\bibfield  {title} {\bibinfo {title} {Non-abelian topological orders and
  majorana fermions in spin-singlet superconductors},\ }\href
  {https://doi.org/10.1103/PhysRevB.82.134521} {\bibfield  {journal} {\bibinfo
  {journal} {Phys. Rev. B}\ }\textbf {\bibinfo {volume} {82}},\ \bibinfo
  {pages} {134521} (\bibinfo {year} {2010})}\BibitemShut {NoStop}%
\bibitem [{\citenamefont {Lutchyn}\ \emph {et~al.}(2010)\citenamefont
  {Lutchyn}, \citenamefont {Sau},\ and\ \citenamefont
  {Das~Sarma}}]{S_SC_Heterostructures}%
  \BibitemOpen
  \bibfield  {author} {\bibinfo {author} {\bibfnamefont {R.~M.}\ \bibnamefont
  {Lutchyn}}, \bibinfo {author} {\bibfnamefont {J.~D.}\ \bibnamefont {Sau}},\
  and\ \bibinfo {author} {\bibfnamefont {S.}~\bibnamefont {Das~Sarma}},\
  }\bibfield  {title} {\bibinfo {title} {Majorana fermions and a topological
  phase transition in semiconductor-superconductor heterostructures},\ }\href
  {https://doi.org/10.1103/PhysRevLett.105.077001} {\bibfield  {journal}
  {\bibinfo  {journal} {Phys. Rev. Lett.}\ }\textbf {\bibinfo {volume} {105}},\
  \bibinfo {pages} {077001} (\bibinfo {year} {2010})}\BibitemShut {NoStop}%
\bibitem [{\citenamefont {Mourik}\ \emph {et~al.}(2012)\citenamefont {Mourik},
  \citenamefont {Zuo}, \citenamefont {Frolov}, \citenamefont {Plissard},
  \citenamefont {Bakkers},\ and\ \citenamefont
  {Kouwenhoven}}]{Signatures_of_MF}%
  \BibitemOpen
  \bibfield  {author} {\bibinfo {author} {\bibfnamefont {V.}~\bibnamefont
  {Mourik}}, \bibinfo {author} {\bibfnamefont {K.}~\bibnamefont {Zuo}},
  \bibinfo {author} {\bibfnamefont {S.~M.}\ \bibnamefont {Frolov}}, \bibinfo
  {author} {\bibfnamefont {S.~R.}\ \bibnamefont {Plissard}}, \bibinfo {author}
  {\bibfnamefont {E.~P. A.~M.}\ \bibnamefont {Bakkers}},\ and\ \bibinfo
  {author} {\bibfnamefont {L.~P.}\ \bibnamefont {Kouwenhoven}},\ }\bibfield
  {title} {\bibinfo {title} {Signatures of majorana fermions in hybrid
  superconductor-semiconductor nanowire devices},\ }\href
  {https://doi.org/10.1126/science.1222360} {\bibfield  {journal} {\bibinfo
  {journal} {Science}\ }\textbf {\bibinfo {volume} {336}},\ \bibinfo {pages}
  {1003} (\bibinfo {year} {2012})}\BibitemShut {NoStop}%
\bibitem [{\citenamefont {Das}\ \emph {et~al.}(2012)\citenamefont {Das},
  \citenamefont {Ronen}, \citenamefont {Most}, \citenamefont {Oreg},
  \citenamefont {Heiblum},\ and\ \citenamefont {Shtrikman}}]{Zero_bias_peaks}%
  \BibitemOpen
  \bibfield  {author} {\bibinfo {author} {\bibfnamefont {A.}~\bibnamefont
  {Das}}, \bibinfo {author} {\bibfnamefont {Y.}~\bibnamefont {Ronen}}, \bibinfo
  {author} {\bibfnamefont {Y.}~\bibnamefont {Most}}, \bibinfo {author}
  {\bibfnamefont {Y.}~\bibnamefont {Oreg}}, \bibinfo {author} {\bibfnamefont
  {M.}~\bibnamefont {Heiblum}},\ and\ \bibinfo {author} {\bibfnamefont
  {H.}~\bibnamefont {Shtrikman}},\ }\bibfield  {title} {\bibinfo {title}
  {Zero-bias peaks and splitting in an al-inas nanowire topological
  superconductor as a signature of majorana fermions},\ }\href
  {https://doi.org/https://doi.org/10.1038/nphys2479} {\bibfield  {journal}
  {\bibinfo  {journal} {Nature Physics}\ }\textbf {\bibinfo {volume} {8}},\
  \bibinfo {pages} {887} (\bibinfo {year} {2012})}\BibitemShut {NoStop}%
\bibitem [{\citenamefont {Rokhinson}\ \emph {et~al.}(2012)\citenamefont
  {Rokhinson}, \citenamefont {Liu},\ and\ \citenamefont
  {Furdyna}}]{ac_Josephson}%
  \BibitemOpen
  \bibfield  {author} {\bibinfo {author} {\bibfnamefont {L.~P.}\ \bibnamefont
  {Rokhinson}}, \bibinfo {author} {\bibfnamefont {X.}~\bibnamefont {Liu}},\
  and\ \bibinfo {author} {\bibfnamefont {J.~K.}\ \bibnamefont {Furdyna}},\
  }\bibfield  {title} {\bibinfo {title} {The fractional a.c. josephson effect
  in a semiconductor-superconductor nanowire as signature of majorana
  particles},\ }\href {https://doi.org/https://doi.org/10.1038/nphys2429}
  {\bibfield  {journal} {\bibinfo  {journal} {Nature Physics}\ }\textbf
  {\bibinfo {volume} {8}},\ \bibinfo {pages} {795} (\bibinfo {year}
  {2012})}\BibitemShut {NoStop}%
\bibitem [{\citenamefont {Deng}\ \emph {et~al.}(2012)\citenamefont {Deng},
  \citenamefont {Yu}, \citenamefont {Huang}, \citenamefont {Larsson},
  \citenamefont {Caroff},\ and\ \citenamefont
  {Xu}}]{Anomalous_Zero_Bias_Conductance}%
  \BibitemOpen
  \bibfield  {author} {\bibinfo {author} {\bibfnamefont {M.~T.}\ \bibnamefont
  {Deng}}, \bibinfo {author} {\bibfnamefont {C.~L.}\ \bibnamefont {Yu}},
  \bibinfo {author} {\bibfnamefont {G.~Y.}\ \bibnamefont {Huang}}, \bibinfo
  {author} {\bibfnamefont {M.}~\bibnamefont {Larsson}}, \bibinfo {author}
  {\bibfnamefont {P.}~\bibnamefont {Caroff}},\ and\ \bibinfo {author}
  {\bibfnamefont {H.~Q.}\ \bibnamefont {Xu}},\ }\bibfield  {title} {\bibinfo
  {title} {Anomalous zero-bias conductance peak in a nb-insb nanowire-nb hybrid
  device},\ }\href {https://doi.org/10.1021/nl303758w} {\bibfield  {journal}
  {\bibinfo  {journal} {Nano Letters}\ }\textbf {\bibinfo {volume} {12}},\
  \bibinfo {pages} {6414} (\bibinfo {year} {2012})}\BibitemShut {NoStop}%
\bibitem [{\citenamefont {Tewari}\ \emph {et~al.}(2007)\citenamefont {Tewari},
  \citenamefont {Das~Sarma}, \citenamefont {Nayak}, \citenamefont {Zhang},\
  and\ \citenamefont {Zoller}}]{MZM_Fermionic_Cold_Atoms}%
  \BibitemOpen
  \bibfield  {author} {\bibinfo {author} {\bibfnamefont {S.}~\bibnamefont
  {Tewari}}, \bibinfo {author} {\bibfnamefont {S.}~\bibnamefont {Das~Sarma}},
  \bibinfo {author} {\bibfnamefont {C.}~\bibnamefont {Nayak}}, \bibinfo
  {author} {\bibfnamefont {C.}~\bibnamefont {Zhang}},\ and\ \bibinfo {author}
  {\bibfnamefont {P.}~\bibnamefont {Zoller}},\ }\bibfield  {title} {\bibinfo
  {title} {Quantum computation using vortices and majorana zero modes of a
  ${p}_{x}+i{p}_{y}$ superfluid of fermionic cold atoms},\ }\href
  {https://doi.org/10.1103/PhysRevLett.98.010506} {\bibfield  {journal}
  {\bibinfo  {journal} {Phys. Rev. Lett.}\ }\textbf {\bibinfo {volume} {98}},\
  \bibinfo {pages} {010506} (\bibinfo {year} {2007})}\BibitemShut {NoStop}%
\bibitem [{\citenamefont {Shermadini}\ \emph {et~al.}(2011)\citenamefont
  {Shermadini}, \citenamefont {Krzton-Maziopa}, \citenamefont {Bendele},
  \citenamefont {Khasanov}, \citenamefont {Luetkens}, \citenamefont {Conder},
  \citenamefont {Pomjakushina}, \citenamefont {Weyeneth}, \citenamefont
  {Pomjakushin}, \citenamefont {Bossen},\ and\ \citenamefont
  {Amato}}]{MF_Cold_Atom_Quantum_Wires}%
  \BibitemOpen
  \bibfield  {author} {\bibinfo {author} {\bibfnamefont {Z.}~\bibnamefont
  {Shermadini}}, \bibinfo {author} {\bibfnamefont {A.}~\bibnamefont
  {Krzton-Maziopa}}, \bibinfo {author} {\bibfnamefont {M.}~\bibnamefont
  {Bendele}}, \bibinfo {author} {\bibfnamefont {R.}~\bibnamefont {Khasanov}},
  \bibinfo {author} {\bibfnamefont {H.}~\bibnamefont {Luetkens}}, \bibinfo
  {author} {\bibfnamefont {K.}~\bibnamefont {Conder}}, \bibinfo {author}
  {\bibfnamefont {E.}~\bibnamefont {Pomjakushina}}, \bibinfo {author}
  {\bibfnamefont {S.}~\bibnamefont {Weyeneth}}, \bibinfo {author}
  {\bibfnamefont {V.}~\bibnamefont {Pomjakushin}}, \bibinfo {author}
  {\bibfnamefont {O.}~\bibnamefont {Bossen}},\ and\ \bibinfo {author}
  {\bibfnamefont {A.}~\bibnamefont {Amato}},\ }\bibfield  {title} {\bibinfo
  {title} {Coexistence of magnetism and superconductivity in the iron-based
  compound ${\mathrm{cs}}_{0.8}({\mathrm{fese}}_{0.98}{)}_{2}$},\ }\href
  {https://doi.org/10.1103/PhysRevLett.106.117602} {\bibfield  {journal}
  {\bibinfo  {journal} {Phys. Rev. Lett.}\ }\textbf {\bibinfo {volume} {106}},\
  \bibinfo {pages} {117602} (\bibinfo {year} {2011})}\BibitemShut {NoStop}%
\bibitem [{\citenamefont {Liu}\ \emph {et~al.}(2014)\citenamefont {Liu},
  \citenamefont {Law},\ and\ \citenamefont {Ng}}]{2D_Spin_Orbit}%
  \BibitemOpen
  \bibfield  {author} {\bibinfo {author} {\bibfnamefont {X.-J.}\ \bibnamefont
  {Liu}}, \bibinfo {author} {\bibfnamefont {K.~T.}\ \bibnamefont {Law}},\ and\
  \bibinfo {author} {\bibfnamefont {T.~K.}\ \bibnamefont {Ng}},\ }\bibfield
  {title} {\bibinfo {title} {Realization of 2d spin-orbit interaction and
  exotic topological orders in cold atoms},\ }\href
  {https://doi.org/10.1103/PhysRevLett.112.086401} {\bibfield  {journal}
  {\bibinfo  {journal} {Phys. Rev. Lett.}\ }\textbf {\bibinfo {volume} {112}},\
  \bibinfo {pages} {086401} (\bibinfo {year} {2014})}\BibitemShut {NoStop}%
\bibitem [{\citenamefont {Sato}\ \emph {et~al.}(2009)\citenamefont {Sato},
  \citenamefont {Takahashi},\ and\ \citenamefont {Fujimoto}}]{s_Wave_SF}%
  \BibitemOpen
  \bibfield  {author} {\bibinfo {author} {\bibfnamefont {M.}~\bibnamefont
  {Sato}}, \bibinfo {author} {\bibfnamefont {Y.}~\bibnamefont {Takahashi}},\
  and\ \bibinfo {author} {\bibfnamefont {S.}~\bibnamefont {Fujimoto}},\
  }\bibfield  {title} {\bibinfo {title} {Non-abelian topological order in
  $s$-wave superfluids of ultracold fermionic atoms},\ }\href
  {https://doi.org/10.1103/PhysRevLett.103.020401} {\bibfield  {journal}
  {\bibinfo  {journal} {Phys. Rev. Lett.}\ }\textbf {\bibinfo {volume} {103}},\
  \bibinfo {pages} {020401} (\bibinfo {year} {2009})}\BibitemShut {NoStop}%
\bibitem [{\citenamefont {Wakatsuki}\ \emph
  {et~al.}(2014{\natexlab{a}})\citenamefont {Wakatsuki}, \citenamefont
  {Ezawa},\ and\ \citenamefont {Nagaosa}}]{multichains}%
  \BibitemOpen
  \bibfield  {author} {\bibinfo {author} {\bibfnamefont {R.}~\bibnamefont
  {Wakatsuki}}, \bibinfo {author} {\bibfnamefont {M.}~\bibnamefont {Ezawa}},\
  and\ \bibinfo {author} {\bibfnamefont {N.}~\bibnamefont {Nagaosa}},\
  }\bibfield  {title} {\bibinfo {title} {Majorana fermions and multiple
  topological phase transition in kitaev ladder topological superconductors},\
  }\href {https://doi.org/10.1103/PhysRevB.89.174514} {\bibfield  {journal}
  {\bibinfo  {journal} {Phys. Rev. B}\ }\textbf {\bibinfo {volume} {89}},\
  \bibinfo {pages} {174514} (\bibinfo {year} {2014}{\natexlab{a}})}\BibitemShut
  {NoStop}%
\bibitem [{\citenamefont {Wakatsuki}\ \emph
  {et~al.}(2014{\natexlab{b}})\citenamefont {Wakatsuki}, \citenamefont {Ezawa},
  \citenamefont {Tanaka},\ and\ \citenamefont {Nagaosa}}]{dimerize}%
  \BibitemOpen
  \bibfield  {author} {\bibinfo {author} {\bibfnamefont {R.}~\bibnamefont
  {Wakatsuki}}, \bibinfo {author} {\bibfnamefont {M.}~\bibnamefont {Ezawa}},
  \bibinfo {author} {\bibfnamefont {Y.}~\bibnamefont {Tanaka}},\ and\ \bibinfo
  {author} {\bibfnamefont {N.}~\bibnamefont {Nagaosa}},\ }\bibfield  {title}
  {\bibinfo {title} {Fermion fractionalization to majorana fermions in a
  dimerized kitaev superconductor},\ }\href
  {https://doi.org/10.1103/PhysRevB.90.014505} {\bibfield  {journal} {\bibinfo
  {journal} {Phys. Rev. B}\ }\textbf {\bibinfo {volume} {90}},\ \bibinfo
  {pages} {014505} (\bibinfo {year} {2014}{\natexlab{b}})}\BibitemShut
  {NoStop}%
\bibitem [{\citenamefont {Alecce}\ and\ \citenamefont
  {Dell'Anna}(2017)}]{long_range_hopping_and_pairing}%
  \BibitemOpen
  \bibfield  {author} {\bibinfo {author} {\bibfnamefont {A.}~\bibnamefont
  {Alecce}}\ and\ \bibinfo {author} {\bibfnamefont {L.}~\bibnamefont
  {Dell'Anna}},\ }\bibfield  {title} {\bibinfo {title} {Extended kitaev chain
  with longer-range hopping and pairing},\ }\href
  {https://doi.org/10.1103/PhysRevB.95.195160} {\bibfield  {journal} {\bibinfo
  {journal} {Phys. Rev. B}\ }\textbf {\bibinfo {volume} {95}},\ \bibinfo
  {pages} {195160} (\bibinfo {year} {2017})}\BibitemShut {NoStop}%
\bibitem [{\citenamefont {Vodola}\ \emph {et~al.}(2014)\citenamefont {Vodola},
  \citenamefont {Lepori}, \citenamefont {Ercolessi}, \citenamefont {Gorshkov},\
  and\ \citenamefont {Pupillo}}]{long_range_interaction}%
  \BibitemOpen
  \bibfield  {author} {\bibinfo {author} {\bibfnamefont {D.}~\bibnamefont
  {Vodola}}, \bibinfo {author} {\bibfnamefont {L.}~\bibnamefont {Lepori}},
  \bibinfo {author} {\bibfnamefont {E.}~\bibnamefont {Ercolessi}}, \bibinfo
  {author} {\bibfnamefont {A.~V.}\ \bibnamefont {Gorshkov}},\ and\ \bibinfo
  {author} {\bibfnamefont {G.}~\bibnamefont {Pupillo}},\ }\bibfield  {title}
  {\bibinfo {title} {Kitaev chains with long-range pairing},\ }\href
  {https://doi.org/10.1103/PhysRevLett.113.156402} {\bibfield  {journal}
  {\bibinfo  {journal} {Phys. Rev. Lett.}\ }\textbf {\bibinfo {volume} {113}},\
  \bibinfo {pages} {156402} (\bibinfo {year} {2014})}\BibitemShut {NoStop}%
\bibitem [{\citenamefont {Dudin}\ and\ \citenamefont {Kuzmich}(2012)}]{Dudin}%
  \BibitemOpen
  \bibfield  {author} {\bibinfo {author} {\bibfnamefont {Y.~O.}\ \bibnamefont
  {Dudin}}\ and\ \bibinfo {author} {\bibfnamefont {A.}~\bibnamefont
  {Kuzmich}},\ }\bibfield  {title} {\bibinfo {title} {Strongly interacting
  rydberg excitations of a cold atomic gas},\ }\href
  {https://doi.org/10.1126/science.1217901} {\bibfield  {journal} {\bibinfo
  {journal} {Science}\ }\textbf {\bibinfo {volume} {336}},\ \bibinfo {pages}
  {887} (\bibinfo {year} {2012})}\BibitemShut {NoStop}%
\bibitem [{\citenamefont {Raitzsch}\ \emph {et~al.}(2008)\citenamefont
  {Raitzsch}, \citenamefont {Bendkowsky}, \citenamefont {Heidemann},
  \citenamefont {Butscher}, \citenamefont {L\"ow},\ and\ \citenamefont
  {Pfau}}]{Raitzsch}%
  \BibitemOpen
  \bibfield  {author} {\bibinfo {author} {\bibfnamefont {U.}~\bibnamefont
  {Raitzsch}}, \bibinfo {author} {\bibfnamefont {V.}~\bibnamefont
  {Bendkowsky}}, \bibinfo {author} {\bibfnamefont {R.}~\bibnamefont
  {Heidemann}}, \bibinfo {author} {\bibfnamefont {B.}~\bibnamefont {Butscher}},
  \bibinfo {author} {\bibfnamefont {R.}~\bibnamefont {L\"ow}},\ and\ \bibinfo
  {author} {\bibfnamefont {T.}~\bibnamefont {Pfau}},\ }\bibfield  {title}
  {\bibinfo {title} {Echo experiments in a strongly interacting rydberg gas},\
  }\href {https://doi.org/10.1103/PhysRevLett.100.013002} {\bibfield  {journal}
  {\bibinfo  {journal} {Phys. Rev. Lett.}\ }\textbf {\bibinfo {volume} {100}},\
  \bibinfo {pages} {013002} (\bibinfo {year} {2008})}\BibitemShut {NoStop}%
\bibitem [{\citenamefont {Pritchard}\ \emph {et~al.}(2010)\citenamefont
  {Pritchard}, \citenamefont {Maxwell}, \citenamefont {Gauguet}, \citenamefont
  {Weatherill}, \citenamefont {Jones},\ and\ \citenamefont
  {Adams}}]{Pritchard}%
  \BibitemOpen
  \bibfield  {author} {\bibinfo {author} {\bibfnamefont {J.~D.}\ \bibnamefont
  {Pritchard}}, \bibinfo {author} {\bibfnamefont {D.}~\bibnamefont {Maxwell}},
  \bibinfo {author} {\bibfnamefont {A.}~\bibnamefont {Gauguet}}, \bibinfo
  {author} {\bibfnamefont {K.~J.}\ \bibnamefont {Weatherill}}, \bibinfo
  {author} {\bibfnamefont {M.~P.~A.}\ \bibnamefont {Jones}},\ and\ \bibinfo
  {author} {\bibfnamefont {C.~S.}\ \bibnamefont {Adams}},\ }\bibfield  {title}
  {\bibinfo {title} {Cooperative atom-light interaction in a blockaded rydberg
  ensemble},\ }\href {https://doi.org/10.1103/PhysRevLett.105.193603}
  {\bibfield  {journal} {\bibinfo  {journal} {Phys. Rev. Lett.}\ }\textbf
  {\bibinfo {volume} {105}},\ \bibinfo {pages} {193603} (\bibinfo {year}
  {2010})}\BibitemShut {NoStop}%
\bibitem [{\citenamefont {Nipper}\ \emph {et~al.}(2012)\citenamefont {Nipper},
  \citenamefont {Balewski}, \citenamefont {Krupp}, \citenamefont {Butscher},
  \citenamefont {L\"ow},\ and\ \citenamefont {Pfau}}]{Nipper}%
  \BibitemOpen
  \bibfield  {author} {\bibinfo {author} {\bibfnamefont {J.}~\bibnamefont
  {Nipper}}, \bibinfo {author} {\bibfnamefont {J.~B.}\ \bibnamefont
  {Balewski}}, \bibinfo {author} {\bibfnamefont {A.~T.}\ \bibnamefont {Krupp}},
  \bibinfo {author} {\bibfnamefont {B.}~\bibnamefont {Butscher}}, \bibinfo
  {author} {\bibfnamefont {R.}~\bibnamefont {L\"ow}},\ and\ \bibinfo {author}
  {\bibfnamefont {T.}~\bibnamefont {Pfau}},\ }\bibfield  {title} {\bibinfo
  {title} {Highly resolved measurements of stark-tuned f\"orster resonances
  between rydberg atoms},\ }\href
  {https://doi.org/10.1103/PhysRevLett.108.113001} {\bibfield  {journal}
  {\bibinfo  {journal} {Phys. Rev. Lett.}\ }\textbf {\bibinfo {volume} {108}},\
  \bibinfo {pages} {113001} (\bibinfo {year} {2012})}\BibitemShut {NoStop}%
\bibitem [{\citenamefont {Schau{\ss}}\ \emph {et~al.}(2012)\citenamefont
  {Schau{\ss}}, \citenamefont {Cheneau}, \citenamefont {Endres}, \citenamefont
  {Fukuhara}, \citenamefont {Hild}, \citenamefont {Omran}, \citenamefont
  {Pohl}, \citenamefont {Gross}, \citenamefont {Kuhr},\ and\ \citenamefont
  {Bloch}}]{Schaus}%
  \BibitemOpen
  \bibfield  {author} {\bibinfo {author} {\bibfnamefont {P.}~\bibnamefont
  {Schau{\ss}}}, \bibinfo {author} {\bibfnamefont {M.}~\bibnamefont {Cheneau}},
  \bibinfo {author} {\bibfnamefont {M.}~\bibnamefont {Endres}}, \bibinfo
  {author} {\bibfnamefont {T.}~\bibnamefont {Fukuhara}}, \bibinfo {author}
  {\bibfnamefont {S.}~\bibnamefont {Hild}}, \bibinfo {author} {\bibfnamefont
  {A.}~\bibnamefont {Omran}}, \bibinfo {author} {\bibfnamefont
  {T.}~\bibnamefont {Pohl}}, \bibinfo {author} {\bibfnamefont {C.}~\bibnamefont
  {Gross}}, \bibinfo {author} {\bibfnamefont {S.}~\bibnamefont {Kuhr}},\ and\
  \bibinfo {author} {\bibfnamefont {I.}~\bibnamefont {Bloch}},\ }\bibfield
  {title} {\bibinfo {title} {Observation of spatially ordered structures in a
  two-dimensional rydberg gas},\ }\href {https://doi.org/10.1038/nature11596}
  {\bibfield  {journal} {\bibinfo  {journal} {Nature}\ }\textbf {\bibinfo
  {volume} {491}},\ \bibinfo {pages} {87} (\bibinfo {year} {2012})}\BibitemShut
  {NoStop}%
\bibitem [{\citenamefont {Jaksch}\ \emph {et~al.}(2000)\citenamefont {Jaksch},
  \citenamefont {Cirac}, \citenamefont {Zoller}, \citenamefont {Rolston},
  \citenamefont {C\^ot\'e},\ and\ \citenamefont {Lukin}}]{Jaksch}%
  \BibitemOpen
  \bibfield  {author} {\bibinfo {author} {\bibfnamefont {D.}~\bibnamefont
  {Jaksch}}, \bibinfo {author} {\bibfnamefont {J.~I.}\ \bibnamefont {Cirac}},
  \bibinfo {author} {\bibfnamefont {P.}~\bibnamefont {Zoller}}, \bibinfo
  {author} {\bibfnamefont {S.~L.}\ \bibnamefont {Rolston}}, \bibinfo {author}
  {\bibfnamefont {R.}~\bibnamefont {C\^ot\'e}},\ and\ \bibinfo {author}
  {\bibfnamefont {M.~D.}\ \bibnamefont {Lukin}},\ }\bibfield  {title} {\bibinfo
  {title} {Fast quantum gates for neutral atoms},\ }\href
  {https://doi.org/10.1103/PhysRevLett.85.2208} {\bibfield  {journal} {\bibinfo
   {journal} {Phys. Rev. Lett.}\ }\textbf {\bibinfo {volume} {85}},\ \bibinfo
  {pages} {2208} (\bibinfo {year} {2000})}\BibitemShut {NoStop}%
\bibitem [{\citenamefont {Lukin}\ \emph {et~al.}(2001)\citenamefont {Lukin},
  \citenamefont {Fleischhauer}, \citenamefont {Cote}, \citenamefont {Duan},
  \citenamefont {Jaksch}, \citenamefont {Cirac},\ and\ \citenamefont
  {Zoller}}]{Lukin}%
  \BibitemOpen
  \bibfield  {author} {\bibinfo {author} {\bibfnamefont {M.~D.}\ \bibnamefont
  {Lukin}}, \bibinfo {author} {\bibfnamefont {M.}~\bibnamefont {Fleischhauer}},
  \bibinfo {author} {\bibfnamefont {R.}~\bibnamefont {Cote}}, \bibinfo {author}
  {\bibfnamefont {L.~M.}\ \bibnamefont {Duan}}, \bibinfo {author}
  {\bibfnamefont {D.}~\bibnamefont {Jaksch}}, \bibinfo {author} {\bibfnamefont
  {J.~I.}\ \bibnamefont {Cirac}},\ and\ \bibinfo {author} {\bibfnamefont
  {P.}~\bibnamefont {Zoller}},\ }\bibfield  {title} {\bibinfo {title} {Dipole
  blockade and quantum information processing in mesoscopic atomic ensembles},\
  }\href {https://doi.org/10.1103/PhysRevLett.87.037901} {\bibfield  {journal}
  {\bibinfo  {journal} {Phys. Rev. Lett.}\ }\textbf {\bibinfo {volume} {87}},\
  \bibinfo {pages} {037901} (\bibinfo {year} {2001})}\BibitemShut {NoStop}%
\bibitem [{\citenamefont {Urban}\ \emph {et~al.}(2009)\citenamefont {Urban},
  \citenamefont {Johnson}, \citenamefont {Henage}, \citenamefont {Isenhower},
  \citenamefont {Yavuz}, \citenamefont {Walker},\ and\ \citenamefont
  {Saffman}}]{Urban}%
  \BibitemOpen
  \bibfield  {author} {\bibinfo {author} {\bibfnamefont {E.}~\bibnamefont
  {Urban}}, \bibinfo {author} {\bibfnamefont {T.~A.}\ \bibnamefont {Johnson}},
  \bibinfo {author} {\bibfnamefont {T.}~\bibnamefont {Henage}}, \bibinfo
  {author} {\bibfnamefont {L.}~\bibnamefont {Isenhower}}, \bibinfo {author}
  {\bibfnamefont {D.~D.}\ \bibnamefont {Yavuz}}, \bibinfo {author}
  {\bibfnamefont {T.~G.}\ \bibnamefont {Walker}},\ and\ \bibinfo {author}
  {\bibfnamefont {M.}~\bibnamefont {Saffman}},\ }\bibfield  {title} {\bibinfo
  {title} {Observation of rydberg blockade between two atoms},\ }\href
  {https://doi.org/10.1038/nphys1178} {\bibfield  {journal} {\bibinfo
  {journal} {Nature Physics}\ }\textbf {\bibinfo {volume} {5}},\ \bibinfo
  {pages} {110} (\bibinfo {year} {2009})}\BibitemShut {NoStop}%
\bibitem [{\citenamefont {Ga\"{e}tan}\ \emph {et~al.}(2009)\citenamefont
  {Ga\"{e}tan}, \citenamefont {Miroshnychenko}, \citenamefont {Wilk},
  \citenamefont {Chotia}, \citenamefont {Viteau}, \citenamefont {Comparat},
  \citenamefont {Pillet}, \citenamefont {Browaeys},\ and\ \citenamefont
  {Grangier}}]{Gaetan}%
  \BibitemOpen
  \bibfield  {author} {\bibinfo {author} {\bibfnamefont {A.}~\bibnamefont
  {Ga\"{e}tan}}, \bibinfo {author} {\bibfnamefont {Y.}~\bibnamefont
  {Miroshnychenko}}, \bibinfo {author} {\bibfnamefont {T.}~\bibnamefont
  {Wilk}}, \bibinfo {author} {\bibfnamefont {A.}~\bibnamefont {Chotia}},
  \bibinfo {author} {\bibfnamefont {M.}~\bibnamefont {Viteau}}, \bibinfo
  {author} {\bibfnamefont {D.}~\bibnamefont {Comparat}}, \bibinfo {author}
  {\bibfnamefont {P.}~\bibnamefont {Pillet}}, \bibinfo {author} {\bibfnamefont
  {A.}~\bibnamefont {Browaeys}},\ and\ \bibinfo {author} {\bibfnamefont
  {P.}~\bibnamefont {Grangier}},\ }\bibfield  {title} {\bibinfo {title}
  {Observation of collective excitation of two individual atoms in the rydberg
  blockade regime},\ }\href {https://doi.org/10.1038/nphys1183} {\bibfield
  {journal} {\bibinfo  {journal} {Nature Physics}\ }\textbf {\bibinfo {volume}
  {5}},\ \bibinfo {pages} {115} (\bibinfo {year} {2009})}\BibitemShut {NoStop}%
\bibitem [{\citenamefont {Henkel}\ \emph {et~al.}(2010)\citenamefont {Henkel},
  \citenamefont {Nath},\ and\ \citenamefont {Pohl}}]{Henkel1}%
  \BibitemOpen
  \bibfield  {author} {\bibinfo {author} {\bibfnamefont {N.}~\bibnamefont
  {Henkel}}, \bibinfo {author} {\bibfnamefont {R.}~\bibnamefont {Nath}},\ and\
  \bibinfo {author} {\bibfnamefont {T.}~\bibnamefont {Pohl}},\ }\bibfield
  {title} {\bibinfo {title} {Three-dimensional roton excitations and supersolid
  formation in rydberg-excited bose-einstein condensates},\ }\href
  {https://doi.org/10.1103/PhysRevLett.104.195302} {\bibfield  {journal}
  {\bibinfo  {journal} {Phys. Rev. Lett.}\ }\textbf {\bibinfo {volume} {104}},\
  \bibinfo {pages} {195302} (\bibinfo {year} {2010})}\BibitemShut {NoStop}%
\bibitem [{\citenamefont {Henkel}\ \emph {et~al.}(2012)\citenamefont {Henkel},
  \citenamefont {Cinti}, \citenamefont {Jain}, \citenamefont {Pupillo},\ and\
  \citenamefont {Pohl}}]{Henkel2}%
  \BibitemOpen
  \bibfield  {author} {\bibinfo {author} {\bibfnamefont {N.}~\bibnamefont
  {Henkel}}, \bibinfo {author} {\bibfnamefont {F.}~\bibnamefont {Cinti}},
  \bibinfo {author} {\bibfnamefont {P.}~\bibnamefont {Jain}}, \bibinfo {author}
  {\bibfnamefont {G.}~\bibnamefont {Pupillo}},\ and\ \bibinfo {author}
  {\bibfnamefont {T.}~\bibnamefont {Pohl}},\ }\bibfield  {title} {\bibinfo
  {title} {Supersolid vortex crystals in rydberg-dressed bose-einstein
  condensates},\ }\href {https://doi.org/10.1103/PhysRevLett.108.265301}
  {\bibfield  {journal} {\bibinfo  {journal} {Phys. Rev. Lett.}\ }\textbf
  {\bibinfo {volume} {108}},\ \bibinfo {pages} {265301} (\bibinfo {year}
  {2012})}\BibitemShut {NoStop}%
\bibitem [{\citenamefont {Balewski}\ \emph {et~al.}(2014)\citenamefont
  {Balewski}, \citenamefont {Krupp}, \citenamefont {Gaj}, \citenamefont
  {Hofferberth}, \citenamefont {Löw},\ and\ \citenamefont {Pfau}}]{Balewski}%
  \BibitemOpen
  \bibfield  {author} {\bibinfo {author} {\bibfnamefont {J.~B.}\ \bibnamefont
  {Balewski}}, \bibinfo {author} {\bibfnamefont {A.~T.}\ \bibnamefont {Krupp}},
  \bibinfo {author} {\bibfnamefont {A.}~\bibnamefont {Gaj}}, \bibinfo {author}
  {\bibfnamefont {S.}~\bibnamefont {Hofferberth}}, \bibinfo {author}
  {\bibfnamefont {R.}~\bibnamefont {Löw}},\ and\ \bibinfo {author}
  {\bibfnamefont {T.}~\bibnamefont {Pfau}},\ }\bibfield  {title} {\bibinfo
  {title} {Rydberg dressing: understanding of collective many-body effects and
  implications for experiments},\ }\href
  {https://doi.org/10.1088/1367-2630/16/6/063012} {\bibfield  {journal}
  {\bibinfo  {journal} {New Journal of Physics}\ }\textbf {\bibinfo {volume}
  {16}},\ \bibinfo {pages} {063012} (\bibinfo {year} {2014})}\BibitemShut
  {NoStop}%
\bibitem [{\citenamefont {B\"uchler}\ \emph {et~al.}(2007)\citenamefont
  {B\"uchler}, \citenamefont {Demler}, \citenamefont {Lukin}, \citenamefont
  {Micheli}, \citenamefont {Prokof'ev}, \citenamefont {Pupillo},\ and\
  \citenamefont {Zoller}}]{boson_crystal}%
  \BibitemOpen
  \bibfield  {author} {\bibinfo {author} {\bibfnamefont {H.~P.}\ \bibnamefont
  {B\"uchler}}, \bibinfo {author} {\bibfnamefont {E.}~\bibnamefont {Demler}},
  \bibinfo {author} {\bibfnamefont {M.}~\bibnamefont {Lukin}}, \bibinfo
  {author} {\bibfnamefont {A.}~\bibnamefont {Micheli}}, \bibinfo {author}
  {\bibfnamefont {N.}~\bibnamefont {Prokof'ev}}, \bibinfo {author}
  {\bibfnamefont {G.}~\bibnamefont {Pupillo}},\ and\ \bibinfo {author}
  {\bibfnamefont {P.}~\bibnamefont {Zoller}},\ }\bibfield  {title} {\bibinfo
  {title} {Strongly correlated 2d quantum phases with cold polar molecules:
  Controlling the shape of the interaction potential},\ }\href
  {https://doi.org/10.1103/PhysRevLett.98.060404} {\bibfield  {journal}
  {\bibinfo  {journal} {Phys. Rev. Lett.}\ }\textbf {\bibinfo {volume} {98}},\
  \bibinfo {pages} {060404} (\bibinfo {year} {2007})}\BibitemShut {NoStop}%
\bibitem [{\citenamefont {Cinti}\ \emph {et~al.}(2010)\citenamefont {Cinti},
  \citenamefont {Jain}, \citenamefont {Boninsegni}, \citenamefont {Micheli},
  \citenamefont {Zoller},\ and\ \citenamefont {Pupillo}}]{Cinti}%
  \BibitemOpen
  \bibfield  {author} {\bibinfo {author} {\bibfnamefont {F.}~\bibnamefont
  {Cinti}}, \bibinfo {author} {\bibfnamefont {P.}~\bibnamefont {Jain}},
  \bibinfo {author} {\bibfnamefont {M.}~\bibnamefont {Boninsegni}}, \bibinfo
  {author} {\bibfnamefont {A.}~\bibnamefont {Micheli}}, \bibinfo {author}
  {\bibfnamefont {P.}~\bibnamefont {Zoller}},\ and\ \bibinfo {author}
  {\bibfnamefont {G.}~\bibnamefont {Pupillo}},\ }\bibfield  {title} {\bibinfo
  {title} {Supersolid droplet crystal in a dipole-blockaded gas},\ }\href
  {https://doi.org/10.1103/PhysRevLett.105.135301} {\bibfield  {journal}
  {\bibinfo  {journal} {Phys. Rev. Lett.}\ }\textbf {\bibinfo {volume} {105}},\
  \bibinfo {pages} {135301} (\bibinfo {year} {2010})}\BibitemShut {NoStop}%
\bibitem [{\citenamefont {Pupillo}\ \emph {et~al.}(2010)\citenamefont
  {Pupillo}, \citenamefont {Micheli}, \citenamefont {Boninsegni}, \citenamefont
  {Lesanovsky},\ and\ \citenamefont {Zoller}}]{Pupillo}%
  \BibitemOpen
  \bibfield  {author} {\bibinfo {author} {\bibfnamefont {G.}~\bibnamefont
  {Pupillo}}, \bibinfo {author} {\bibfnamefont {A.}~\bibnamefont {Micheli}},
  \bibinfo {author} {\bibfnamefont {M.}~\bibnamefont {Boninsegni}}, \bibinfo
  {author} {\bibfnamefont {I.}~\bibnamefont {Lesanovsky}},\ and\ \bibinfo
  {author} {\bibfnamefont {P.}~\bibnamefont {Zoller}},\ }\bibfield  {title}
  {\bibinfo {title} {Strongly correlated gases of rydberg-dressed atoms:
  Quantum and classical dynamics},\ }\href
  {https://doi.org/10.1103/PhysRevLett.104.223002} {\bibfield  {journal}
  {\bibinfo  {journal} {Phys. Rev. Lett.}\ }\textbf {\bibinfo {volume} {104}},\
  \bibinfo {pages} {223002} (\bibinfo {year} {2010})}\BibitemShut {NoStop}%
\bibitem [{\citenamefont {Maucher}\ \emph {et~al.}(2011)\citenamefont
  {Maucher}, \citenamefont {Henkel}, \citenamefont {Saffman}, \citenamefont
  {Kr\'olikowski}, \citenamefont {Skupin},\ and\ \citenamefont
  {Pohl}}]{Maucher}%
  \BibitemOpen
  \bibfield  {author} {\bibinfo {author} {\bibfnamefont {F.}~\bibnamefont
  {Maucher}}, \bibinfo {author} {\bibfnamefont {N.}~\bibnamefont {Henkel}},
  \bibinfo {author} {\bibfnamefont {M.}~\bibnamefont {Saffman}}, \bibinfo
  {author} {\bibfnamefont {W.}~\bibnamefont {Kr\'olikowski}}, \bibinfo {author}
  {\bibfnamefont {S.}~\bibnamefont {Skupin}},\ and\ \bibinfo {author}
  {\bibfnamefont {T.}~\bibnamefont {Pohl}},\ }\bibfield  {title} {\bibinfo
  {title} {Rydberg-induced solitons: Three-dimensional self-trapping of matter
  waves},\ }\href {https://doi.org/10.1103/PhysRevLett.106.170401} {\bibfield
  {journal} {\bibinfo  {journal} {Phys. Rev. Lett.}\ }\textbf {\bibinfo
  {volume} {106}},\ \bibinfo {pages} {170401} (\bibinfo {year}
  {2011})}\BibitemShut {NoStop}%
\bibitem [{\citenamefont {Xiong}\ \emph {et~al.}(2014)\citenamefont {Xiong},
  \citenamefont {Jen},\ and\ \citenamefont {Wang}}]{Xiong}%
  \BibitemOpen
  \bibfield  {author} {\bibinfo {author} {\bibfnamefont {B.}~\bibnamefont
  {Xiong}}, \bibinfo {author} {\bibfnamefont {H.~H.}\ \bibnamefont {Jen}},\
  and\ \bibinfo {author} {\bibfnamefont {D.-W.}\ \bibnamefont {Wang}},\
  }\bibfield  {title} {\bibinfo {title} {Topological superfluid by blockade
  effects in a rydberg-dressed fermi gas},\ }\href
  {https://doi.org/10.1103/PhysRevA.90.013631} {\bibfield  {journal} {\bibinfo
  {journal} {Phys. Rev. A}\ }\textbf {\bibinfo {volume} {90}},\ \bibinfo
  {pages} {013631} (\bibinfo {year} {2014})}\BibitemShut {NoStop}%
\bibitem [{\citenamefont {Li}\ and\ \citenamefont {Sarma}(2015)}]{Li}%
  \BibitemOpen
  \bibfield  {author} {\bibinfo {author} {\bibfnamefont {X.}~\bibnamefont
  {Li}}\ and\ \bibinfo {author} {\bibfnamefont {S.~D.}\ \bibnamefont {Sarma}},\
  }\bibfield  {title} {\bibinfo {title} {Exotic topological density waves in
  cold atomic rydberg-dressed fermions},\ }\href
  {https://doi.org/10.1038/ncomms8137} {\bibfield  {journal} {\bibinfo
  {journal} {Nature Communications}\ }\textbf {\bibinfo {volume} {6}},\
  \bibinfo {pages} {7137} (\bibinfo {year} {2015})}\BibitemShut {NoStop}%
\bibitem [{\citenamefont {Jau}\ \emph {et~al.}(2016)\citenamefont {Jau},
  \citenamefont {Hankin}, \citenamefont {Keating}, \citenamefont {Deutsch},\
  and\ \citenamefont {Biedermann}}]{Jau}%
  \BibitemOpen
  \bibfield  {author} {\bibinfo {author} {\bibfnamefont {Y.-Y.}\ \bibnamefont
  {Jau}}, \bibinfo {author} {\bibfnamefont {A.~M.}\ \bibnamefont {Hankin}},
  \bibinfo {author} {\bibfnamefont {T.}~\bibnamefont {Keating}}, \bibinfo
  {author} {\bibfnamefont {I.~H.}\ \bibnamefont {Deutsch}},\ and\ \bibinfo
  {author} {\bibfnamefont {G.~W.}\ \bibnamefont {Biedermann}},\ }\bibfield
  {title} {\bibinfo {title} {Entangling atomic spins with a rydberg-dressed
  spin-flip blockade},\ }\href {https://doi.org/10.1038/nphys3487} {\bibfield
  {journal} {\bibinfo  {journal} {Nature Physics}\ }\textbf {\bibinfo {volume}
  {12}},\ \bibinfo {pages} {71} (\bibinfo {year} {2016})}\BibitemShut {NoStop}%
\bibitem [{\citenamefont {Zeiher}\ \emph {et~al.}(2013)\citenamefont {Zeiher},
  \citenamefont {van Bijnen}, \citenamefont {Hild}, \citenamefont {Jae-yoon
  Choi}, \citenamefont {Bloch},\ and\ \citenamefont {Gross}}]{Zeiher}%
  \BibitemOpen
  \bibfield  {author} {\bibinfo {author} {\bibfnamefont {J.}~\bibnamefont
  {Zeiher}}, \bibinfo {author} {\bibfnamefont {R.}~\bibnamefont {van Bijnen}},
  \bibinfo {author} {\bibfnamefont {P.~S.~S.}\ \bibnamefont {Hild}}, \bibinfo
  {author} {\bibfnamefont {T.~P.}\ \bibnamefont {Jae-yoon Choi}}, \bibinfo
  {author} {\bibfnamefont {I.}~\bibnamefont {Bloch}},\ and\ \bibinfo {author}
  {\bibfnamefont {C.}~\bibnamefont {Gross}},\ }\bibfield  {title} {\bibinfo
  {title} {s},\ }\href@noop {} {\bibfield  {journal} {\bibinfo  {journal}
  {arXiv: 1602.06313}\ } (\bibinfo {year} {2013})}\BibitemShut {NoStop}%
\bibitem [{\citenamefont {Murakami}(2007)}]{Shuichi_Murakami_2007}%
  \BibitemOpen
  \bibfield  {author} {\bibinfo {author} {\bibfnamefont {S.}~\bibnamefont
  {Murakami}},\ }\bibfield  {title} {\bibinfo {title} {Phase transition between
  the quantum spin hall and insulator phases in 3d: emergence of a topological
  gapless phase},\ }\href {https://doi.org/10.1088/1367-2630/9/9/356}
  {\bibfield  {journal} {\bibinfo  {journal} {New Journal of Physics}\ }\textbf
  {\bibinfo {volume} {9}},\ \bibinfo {pages} {356} (\bibinfo {year}
  {2007})}\BibitemShut {NoStop}%
\bibitem [{\citenamefont {Honer}\ \emph {et~al.}(2010)\citenamefont {Honer},
  \citenamefont {Weimer}, \citenamefont {Pfau},\ and\ \citenamefont
  {B\"uchler}}]{Honer}%
  \BibitemOpen
  \bibfield  {author} {\bibinfo {author} {\bibfnamefont {J.}~\bibnamefont
  {Honer}}, \bibinfo {author} {\bibfnamefont {H.}~\bibnamefont {Weimer}},
  \bibinfo {author} {\bibfnamefont {T.}~\bibnamefont {Pfau}},\ and\ \bibinfo
  {author} {\bibfnamefont {H.~P.}\ \bibnamefont {B\"uchler}},\ }\bibfield
  {title} {\bibinfo {title} {Collective many-body interaction in rydberg
  dressed atoms},\ }\href {https://doi.org/10.1103/PhysRevLett.105.160404}
  {\bibfield  {journal} {\bibinfo  {journal} {Phys. Rev. Lett.}\ }\textbf
  {\bibinfo {volume} {105}},\ \bibinfo {pages} {160404} (\bibinfo {year}
  {2010})}\BibitemShut {NoStop}%
\bibitem [{\citenamefont {Walker}\ and\ \citenamefont
  {Saffman}(2008)}]{Walker}%
  \BibitemOpen
  \bibfield  {author} {\bibinfo {author} {\bibfnamefont {T.~G.}\ \bibnamefont
  {Walker}}\ and\ \bibinfo {author} {\bibfnamefont {M.}~\bibnamefont
  {Saffman}},\ }\bibfield  {title} {\bibinfo {title} {Consequences of zeeman
  degeneracy for the van der waals blockade between rydberg atoms},\ }\href
  {https://doi.org/10.1103/PhysRevA.77.032723} {\bibfield  {journal} {\bibinfo
  {journal} {Phys. Rev. A}\ }\textbf {\bibinfo {volume} {77}},\ \bibinfo
  {pages} {032723} (\bibinfo {year} {2008})}\BibitemShut {NoStop}%
\bibitem [{\citenamefont {Singer}\ \emph {et~al.}(2005)\citenamefont {Singer},
  \citenamefont {Reetz-Lamour}, \citenamefont {Amthor}, \citenamefont
  {Fölling}, \citenamefont {Tscherneck},\ and\ \citenamefont
  {Weidemüller}}]{Singer2}%
  \BibitemOpen
  \bibfield  {author} {\bibinfo {author} {\bibfnamefont {K.}~\bibnamefont
  {Singer}}, \bibinfo {author} {\bibfnamefont {M.}~\bibnamefont
  {Reetz-Lamour}}, \bibinfo {author} {\bibfnamefont {T.}~\bibnamefont
  {Amthor}}, \bibinfo {author} {\bibfnamefont {S.}~\bibnamefont {Fölling}},
  \bibinfo {author} {\bibfnamefont {M.}~\bibnamefont {Tscherneck}},\ and\
  \bibinfo {author} {\bibfnamefont {M.}~\bibnamefont {Weidemüller}},\
  }\bibfield  {title} {\bibinfo {title} {Spectroscopy of an ultracold rydberg
  gas and signatures of rydberg{\textendash}rydberg interactions},\ }\href
  {https://doi.org/10.1088/0953-4075/38/2/023} {\bibfield  {journal} {\bibinfo
  {journal} {Journal of Physics B: Atomic, Molecular and Optical Physics}\
  }\textbf {\bibinfo {volume} {38}},\ \bibinfo {pages} {S321} (\bibinfo {year}
  {2005})}\BibitemShut {NoStop}%
\bibitem [{\citenamefont {Tanaka}\ \emph {et~al.}(2010)\citenamefont {Tanaka},
  \citenamefont {Mizuno}, \citenamefont {Yokoyama}, \citenamefont {Yada},\ and\
  \citenamefont {Sato}}]{PhysRevLett.105.097002}%
  \BibitemOpen
  \bibfield  {author} {\bibinfo {author} {\bibfnamefont {Y.}~\bibnamefont
  {Tanaka}}, \bibinfo {author} {\bibfnamefont {Y.}~\bibnamefont {Mizuno}},
  \bibinfo {author} {\bibfnamefont {T.}~\bibnamefont {Yokoyama}}, \bibinfo
  {author} {\bibfnamefont {K.}~\bibnamefont {Yada}},\ and\ \bibinfo {author}
  {\bibfnamefont {M.}~\bibnamefont {Sato}},\ }\bibfield  {title} {\bibinfo
  {title} {Anomalous andreev bound state in noncentrosymmetric
  superconductors},\ }\href {https://doi.org/10.1103/PhysRevLett.105.097002}
  {\bibfield  {journal} {\bibinfo  {journal} {Phys. Rev. Lett.}\ }\textbf
  {\bibinfo {volume} {105}},\ \bibinfo {pages} {097002} (\bibinfo {year}
  {2010})}\BibitemShut {NoStop}%
\bibitem [{\citenamefont {Sato}\ and\ \citenamefont
  {Fujimoto}(2010)}]{PhysRevLett.105.217001}%
  \BibitemOpen
  \bibfield  {author} {\bibinfo {author} {\bibfnamefont {M.}~\bibnamefont
  {Sato}}\ and\ \bibinfo {author} {\bibfnamefont {S.}~\bibnamefont
  {Fujimoto}},\ }\bibfield  {title} {\bibinfo {title} {Existence of majorana
  fermions and topological order in nodal superconductors with spin-orbit
  interactions in external magnetic fields},\ }\href
  {https://doi.org/10.1103/PhysRevLett.105.217001} {\bibfield  {journal}
  {\bibinfo  {journal} {Phys. Rev. Lett.}\ }\textbf {\bibinfo {volume} {105}},\
  \bibinfo {pages} {217001} (\bibinfo {year} {2010})}\BibitemShut {NoStop}%
\bibitem [{\citenamefont {Schnyder}\ and\ \citenamefont
  {Ryu}(2011)}]{PhysRevB.84.060504}%
  \BibitemOpen
  \bibfield  {author} {\bibinfo {author} {\bibfnamefont {A.~P.}\ \bibnamefont
  {Schnyder}}\ and\ \bibinfo {author} {\bibfnamefont {S.}~\bibnamefont {Ryu}},\
  }\bibfield  {title} {\bibinfo {title} {Topological phases and surface flat
  bands in superconductors without inversion symmetry},\ }\href
  {https://doi.org/10.1103/PhysRevB.84.060504} {\bibfield  {journal} {\bibinfo
  {journal} {Phys. Rev. B}\ }\textbf {\bibinfo {volume} {84}},\ \bibinfo
  {pages} {060504} (\bibinfo {year} {2011})}\BibitemShut {NoStop}%
\bibitem [{\citenamefont {Kitaev}\ and\ \citenamefont
  {Preskill}(2006)}]{EN_TO_1}%
  \BibitemOpen
  \bibfield  {author} {\bibinfo {author} {\bibfnamefont {A.}~\bibnamefont
  {Kitaev}}\ and\ \bibinfo {author} {\bibfnamefont {J.}~\bibnamefont
  {Preskill}},\ }\bibfield  {title} {\bibinfo {title} {Topological entanglement
  entropy},\ }\href {https://doi.org/10.1103/PhysRevLett.96.110404} {\bibfield
  {journal} {\bibinfo  {journal} {Phys. Rev. Lett.}\ }\textbf {\bibinfo
  {volume} {96}},\ \bibinfo {pages} {110404} (\bibinfo {year}
  {2006})}\BibitemShut {NoStop}%
\bibitem [{\citenamefont {Levin}\ and\ \citenamefont {Wen}(2006)}]{EN_TO_2}%
  \BibitemOpen
  \bibfield  {author} {\bibinfo {author} {\bibfnamefont {M.}~\bibnamefont
  {Levin}}\ and\ \bibinfo {author} {\bibfnamefont {X.-G.}\ \bibnamefont
  {Wen}},\ }\bibfield  {title} {\bibinfo {title} {Detecting topological order
  in a ground state wave function},\ }\href
  {https://doi.org/10.1103/PhysRevLett.96.110405} {\bibfield  {journal}
  {\bibinfo  {journal} {Phys. Rev. Lett.}\ }\textbf {\bibinfo {volume} {96}},\
  \bibinfo {pages} {110405} (\bibinfo {year} {2006})}\BibitemShut {NoStop}%
\bibitem [{\citenamefont {Li}\ and\ \citenamefont {Haldane}(2008)}]{ES_FQHE}%
  \BibitemOpen
  \bibfield  {author} {\bibinfo {author} {\bibfnamefont {H.}~\bibnamefont
  {Li}}\ and\ \bibinfo {author} {\bibfnamefont {F.~D.~M.}\ \bibnamefont
  {Haldane}},\ }\bibfield  {title} {\bibinfo {title} {Entanglement spectrum as
  a generalization of entanglement entropy: Identification of topological order
  in non-abelian fractional quantum hall effect states},\ }\href
  {https://doi.org/10.1103/PhysRevLett.101.010504} {\bibfield  {journal}
  {\bibinfo  {journal} {Phys. Rev. Lett.}\ }\textbf {\bibinfo {volume} {101}},\
  \bibinfo {pages} {010504} (\bibinfo {year} {2008})}\BibitemShut {NoStop}%
\bibitem [{\citenamefont {Pollmann}\ \emph {et~al.}(2010)\citenamefont
  {Pollmann}, \citenamefont {Turner}, \citenamefont {Berg},\ and\ \citenamefont
  {Oshikawa}}]{ES_SPT_1}%
  \BibitemOpen
  \bibfield  {author} {\bibinfo {author} {\bibfnamefont {F.}~\bibnamefont
  {Pollmann}}, \bibinfo {author} {\bibfnamefont {A.~M.}\ \bibnamefont
  {Turner}}, \bibinfo {author} {\bibfnamefont {E.}~\bibnamefont {Berg}},\ and\
  \bibinfo {author} {\bibfnamefont {M.}~\bibnamefont {Oshikawa}},\ }\bibfield
  {title} {\bibinfo {title} {Entanglement spectrum of a topological phase in
  one dimension},\ }\href {https://doi.org/10.1103/PhysRevB.81.064439}
  {\bibfield  {journal} {\bibinfo  {journal} {Phys. Rev. B}\ }\textbf {\bibinfo
  {volume} {81}},\ \bibinfo {pages} {064439} (\bibinfo {year}
  {2010})}\BibitemShut {NoStop}%
\bibitem [{\citenamefont {Turner}\ \emph {et~al.}(2011)\citenamefont {Turner},
  \citenamefont {Pollmann},\ and\ \citenamefont {Berg}}]{ES_SPT_2}%
  \BibitemOpen
  \bibfield  {author} {\bibinfo {author} {\bibfnamefont {A.~M.}\ \bibnamefont
  {Turner}}, \bibinfo {author} {\bibfnamefont {F.}~\bibnamefont {Pollmann}},\
  and\ \bibinfo {author} {\bibfnamefont {E.}~\bibnamefont {Berg}},\ }\bibfield
  {title} {\bibinfo {title} {Topological phases of one-dimensional fermions: An
  entanglement point of view},\ }\href
  {https://doi.org/10.1103/PhysRevB.83.075102} {\bibfield  {journal} {\bibinfo
  {journal} {Phys. Rev. B}\ }\textbf {\bibinfo {volume} {83}},\ \bibinfo
  {pages} {075102} (\bibinfo {year} {2011})}\BibitemShut {NoStop}%
\bibitem [{\citenamefont {Chang}\ \emph
  {et~al.}(2014{\natexlab{a}})\citenamefont {Chang}, \citenamefont {Mudry},\
  and\ \citenamefont {Ryu}}]{Chang_2014}%
  \BibitemOpen
  \bibfield  {author} {\bibinfo {author} {\bibfnamefont {P.-Y.}\ \bibnamefont
  {Chang}}, \bibinfo {author} {\bibfnamefont {C.}~\bibnamefont {Mudry}},\ and\
  \bibinfo {author} {\bibfnamefont {S.}~\bibnamefont {Ryu}},\ }\bibfield
  {title} {\bibinfo {title} {Symmetry-protected entangling boundary zero modes
  in crystalline topological insulators},\ }\href
  {https://doi.org/10.1088/1742-5468/2014/09/p09014} {\bibfield  {journal}
  {\bibinfo  {journal} {Journal of Statistical Mechanics: Theory and
  Experiment}\ }\textbf {\bibinfo {volume} {2014}},\ \bibinfo {pages} {P09014}
  (\bibinfo {year} {2014}{\natexlab{a}})}\BibitemShut {NoStop}%
\bibitem [{\citenamefont {Ryu}\ and\ \citenamefont
  {Hatsugai}(2006)}]{berry_phase}%
  \BibitemOpen
  \bibfield  {author} {\bibinfo {author} {\bibfnamefont {S.}~\bibnamefont
  {Ryu}}\ and\ \bibinfo {author} {\bibfnamefont {Y.}~\bibnamefont {Hatsugai}},\
  }\bibfield  {title} {\bibinfo {title} {Entanglement entropy and the berry
  phase in the solid state},\ }\href
  {https://doi.org/10.1103/PhysRevB.73.245115} {\bibfield  {journal} {\bibinfo
  {journal} {Phys. Rev. B}\ }\textbf {\bibinfo {volume} {73}},\ \bibinfo
  {pages} {245115} (\bibinfo {year} {2006})}\BibitemShut {NoStop}%
\bibitem [{\citenamefont {Chung}\ \emph {et~al.}(2011)\citenamefont {Chung},
  \citenamefont {Jhu}, \citenamefont {Chen},\ and\ \citenamefont
  {Yip}}]{opes_1}%
  \BibitemOpen
  \bibfield  {author} {\bibinfo {author} {\bibfnamefont {M.-C.}\ \bibnamefont
  {Chung}}, \bibinfo {author} {\bibfnamefont {Y.-H.}\ \bibnamefont {Jhu}},
  \bibinfo {author} {\bibfnamefont {P.}~\bibnamefont {Chen}},\ and\ \bibinfo
  {author} {\bibfnamefont {S.}~\bibnamefont {Yip}},\ }\bibfield  {title}
  {\bibinfo {title} {Edge states, entanglement entropy spectra and critical
  hopping couplings of anisotropic honeycomb lattices},\ }\href
  {http://stacks.iop.org/0295-5075/95/i=2/a=27003} {\bibfield  {journal}
  {\bibinfo  {journal} {EPL (Europhysics Letters)}\ }\textbf {\bibinfo {volume}
  {95}},\ \bibinfo {pages} {27003} (\bibinfo {year} {2011})}\BibitemShut
  {NoStop}%
\bibitem [{\citenamefont {Chung}\ \emph {et~al.}(2013)\citenamefont {Chung},
  \citenamefont {Jhu}, \citenamefont {Chen},\ and\ \citenamefont
  {Mou}}]{opes_2}%
  \BibitemOpen
  \bibfield  {author} {\bibinfo {author} {\bibfnamefont {M.-C.}\ \bibnamefont
  {Chung}}, \bibinfo {author} {\bibfnamefont {Y.-H.}\ \bibnamefont {Jhu}},
  \bibinfo {author} {\bibfnamefont {P.}~\bibnamefont {Chen}},\ and\ \bibinfo
  {author} {\bibfnamefont {C.-Y.}\ \bibnamefont {Mou}},\ }\bibfield  {title}
  {\bibinfo {title} {Quench dynamics of topological maximally entangled
  states},\ }\href {http://stacks.iop.org/0953-8984/25/i=28/a=285601}
  {\bibfield  {journal} {\bibinfo  {journal} {Journal of Physics: Condensed
  Matter}\ }\textbf {\bibinfo {volume} {25}},\ \bibinfo {pages} {285601}
  (\bibinfo {year} {2013})}\BibitemShut {NoStop}%
\bibitem [{\citenamefont {Matsuura}\ \emph {et~al.}(2013)\citenamefont
  {Matsuura}, \citenamefont {Chang}, \citenamefont {Schnyder},\ and\
  \citenamefont {Ryu}}]{Matsuura_2013}%
  \BibitemOpen
  \bibfield  {author} {\bibinfo {author} {\bibfnamefont {S.}~\bibnamefont
  {Matsuura}}, \bibinfo {author} {\bibfnamefont {P.-Y.}\ \bibnamefont {Chang}},
  \bibinfo {author} {\bibfnamefont {A.~P.}\ \bibnamefont {Schnyder}},\ and\
  \bibinfo {author} {\bibfnamefont {S.}~\bibnamefont {Ryu}},\ }\bibfield
  {title} {\bibinfo {title} {Protected boundary states in gapless topological
  phases},\ }\href {https://doi.org/10.1088/1367-2630/15/6/065001} {\bibfield
  {journal} {\bibinfo  {journal} {New Journal of Physics}\ }\textbf {\bibinfo
  {volume} {15}},\ \bibinfo {pages} {065001} (\bibinfo {year}
  {2013})}\BibitemShut {NoStop}%
\bibitem [{\citenamefont {Read}\ and\ \citenamefont {Green}(2000)}]{Read_2000}%
  \BibitemOpen
  \bibfield  {author} {\bibinfo {author} {\bibfnamefont {N.}~\bibnamefont
  {Read}}\ and\ \bibinfo {author} {\bibfnamefont {D.}~\bibnamefont {Green}},\
  }\bibfield  {title} {\bibinfo {title} {Paired states of fermions in two
  dimensions with breaking of parity and time-reversal symmetries and the
  fractional quantum hall effect},\ }\href
  {https://doi.org/10.1103/PhysRevB.61.10267} {\bibfield  {journal} {\bibinfo
  {journal} {Phys. Rev. B}\ }\textbf {\bibinfo {volume} {61}},\ \bibinfo
  {pages} {10267} (\bibinfo {year} {2000})}\BibitemShut {NoStop}%
\bibitem [{\citenamefont {Sumiyoshi}\ and\ \citenamefont
  {Fujimoto}(2013)}]{Sumiyoshi_2013}%
  \BibitemOpen
  \bibfield  {author} {\bibinfo {author} {\bibfnamefont {H.}~\bibnamefont
  {Sumiyoshi}}\ and\ \bibinfo {author} {\bibfnamefont {S.}~\bibnamefont
  {Fujimoto}},\ }\bibfield  {title} {\bibinfo {title} {Quantum thermal hall
  effect in a time-reversal-symmetry-broken topological superconductor in two
  dimensions: Approach from bulk calculations},\ }\href
  {https://doi.org/10.7566/JPSJ.82.023602} {\bibfield  {journal} {\bibinfo
  {journal} {Journal of the Physical Society of Japan}\ }\textbf {\bibinfo
  {volume} {82}},\ \bibinfo {pages} {023602} (\bibinfo {year} {2013})},\
  \Eprint {https://arxiv.org/abs/https://doi.org/10.7566/JPSJ.82.023602}
  {https://doi.org/10.7566/JPSJ.82.023602} \BibitemShut {NoStop}%
\bibitem [{\citenamefont {Nomura}\ \emph {et~al.}(2012)\citenamefont {Nomura},
  \citenamefont {Ryu}, \citenamefont {Furusaki},\ and\ \citenamefont
  {Nagaosa}}]{Nomura_2012}%
  \BibitemOpen
  \bibfield  {author} {\bibinfo {author} {\bibfnamefont {K.}~\bibnamefont
  {Nomura}}, \bibinfo {author} {\bibfnamefont {S.}~\bibnamefont {Ryu}},
  \bibinfo {author} {\bibfnamefont {A.}~\bibnamefont {Furusaki}},\ and\
  \bibinfo {author} {\bibfnamefont {N.}~\bibnamefont {Nagaosa}},\ }\bibfield
  {title} {\bibinfo {title} {Cross-correlated responses of topological
  superconductors and superfluids},\ }\href
  {https://doi.org/10.1103/PhysRevLett.108.026802} {\bibfield  {journal}
  {\bibinfo  {journal} {Phys. Rev. Lett.}\ }\textbf {\bibinfo {volume} {108}},\
  \bibinfo {pages} {026802} (\bibinfo {year} {2012})}\BibitemShut {NoStop}%
\bibitem [{\citenamefont {Chang}\ \emph
  {et~al.}(2014{\natexlab{b}})\citenamefont {Chang}, \citenamefont {Matsuura},
  \citenamefont {Schnyder},\ and\ \citenamefont {Ryu}}]{Chang_2014_2}%
  \BibitemOpen
  \bibfield  {author} {\bibinfo {author} {\bibfnamefont {P.-Y.}\ \bibnamefont
  {Chang}}, \bibinfo {author} {\bibfnamefont {S.}~\bibnamefont {Matsuura}},
  \bibinfo {author} {\bibfnamefont {A.~P.}\ \bibnamefont {Schnyder}},\ and\
  \bibinfo {author} {\bibfnamefont {S.}~\bibnamefont {Ryu}},\ }\bibfield
  {title} {\bibinfo {title} {Majorana vortex-bound states in three-dimensional
  nodal noncentrosymmetric superconductors},\ }\href
  {https://doi.org/10.1103/PhysRevB.90.174504} {\bibfield  {journal} {\bibinfo
  {journal} {Phys. Rev. B}\ }\textbf {\bibinfo {volume} {90}},\ \bibinfo
  {pages} {174504} (\bibinfo {year} {2014}{\natexlab{b}})}\BibitemShut
  {NoStop}%
\bibitem [{\citenamefont {Yin}\ \emph {et~al.}(2015)\citenamefont {Yin},
  \citenamefont {Wu}, \citenamefont {Wang}, \citenamefont {Ye}, \citenamefont
  {Gong}, \citenamefont {Hou}, \citenamefont {Shan}, \citenamefont {Li},
  \citenamefont {Liang}, \citenamefont {Wu}, \citenamefont {Li}, \citenamefont
  {Ting}, \citenamefont {Wang}, \citenamefont {Hu}, \citenamefont {Hor},
  \citenamefont {Ding},\ and\ \citenamefont {Pan}}]{Yin_2015}%
  \BibitemOpen
  \bibfield  {author} {\bibinfo {author} {\bibfnamefont {J.-X.}\ \bibnamefont
  {Yin}}, \bibinfo {author} {\bibfnamefont {Z.}~\bibnamefont {Wu}}, \bibinfo
  {author} {\bibfnamefont {J.-H.}\ \bibnamefont {Wang}}, \bibinfo {author}
  {\bibfnamefont {Z.-Y.}\ \bibnamefont {Ye}}, \bibinfo {author} {\bibfnamefont
  {J.}~\bibnamefont {Gong}}, \bibinfo {author} {\bibfnamefont {X.-Y.}\
  \bibnamefont {Hou}}, \bibinfo {author} {\bibfnamefont {L.}~\bibnamefont
  {Shan}}, \bibinfo {author} {\bibfnamefont {A.}~\bibnamefont {Li}}, \bibinfo
  {author} {\bibfnamefont {X.-J.}\ \bibnamefont {Liang}}, \bibinfo {author}
  {\bibfnamefont {X.-X.}\ \bibnamefont {Wu}}, \bibinfo {author} {\bibfnamefont
  {J.}~\bibnamefont {Li}}, \bibinfo {author} {\bibfnamefont {C.-S.}\
  \bibnamefont {Ting}}, \bibinfo {author} {\bibfnamefont {Z.-Q.}\ \bibnamefont
  {Wang}}, \bibinfo {author} {\bibfnamefont {J.-P.}\ \bibnamefont {Hu}},
  \bibinfo {author} {\bibfnamefont {P.-H.}\ \bibnamefont {Hor}}, \bibinfo
  {author} {\bibfnamefont {H.}~\bibnamefont {Ding}},\ and\ \bibinfo {author}
  {\bibfnamefont {S.~H.}\ \bibnamefont {Pan}},\ }\bibfield  {title} {\bibinfo
  {title} {Observation of a robust zero-energy bound state in iron-based
  superconductor fe(te,se)},\ }\href {https://doi.org/10.1038/nphys3371}
  {\bibfield  {journal} {\bibinfo  {journal} {Nature Physics}\ }\textbf
  {\bibinfo {volume} {11}},\ \bibinfo {pages} {543} (\bibinfo {year}
  {2015})}\BibitemShut {NoStop}%
\bibitem [{\citenamefont {Palacio-Morales}\ \emph {et~al.}(2019)\citenamefont
  {Palacio-Morales}, \citenamefont {Mascot}, \citenamefont {Cocklin},
  \citenamefont {Kim}, \citenamefont {Rachel}, \citenamefont {Morr},\ and\
  \citenamefont {Wiesendanger}}]{Palacio-Moraleseaav6600}%
  \BibitemOpen
  \bibfield  {author} {\bibinfo {author} {\bibfnamefont {A.}~\bibnamefont
  {Palacio-Morales}}, \bibinfo {author} {\bibfnamefont {E.}~\bibnamefont
  {Mascot}}, \bibinfo {author} {\bibfnamefont {S.}~\bibnamefont {Cocklin}},
  \bibinfo {author} {\bibfnamefont {H.}~\bibnamefont {Kim}}, \bibinfo {author}
  {\bibfnamefont {S.}~\bibnamefont {Rachel}}, \bibinfo {author} {\bibfnamefont
  {D.~K.}\ \bibnamefont {Morr}},\ and\ \bibinfo {author} {\bibfnamefont
  {R.}~\bibnamefont {Wiesendanger}},\ }\bibfield  {title} {\bibinfo {title}
  {Atomic-scale interface engineering of majorana edge modes in a 2d
  magnet-superconductor hybrid system},\ }\bibfield  {journal} {\bibinfo
  {journal} {Science Advances}\ }\textbf {\bibinfo {volume} {5}},\ \href
  {https://doi.org/10.1126/sciadv.aav6600} {10.1126/sciadv.aav6600} (\bibinfo
  {year} {2019}),\ \Eprint
  {https://arxiv.org/abs/https://advances.sciencemag.org/content/5/7/eaav6600.full.pdf}
  {https://advances.sciencemag.org/content/5/7/eaav6600.full.pdf} \BibitemShut
  {NoStop}%
\bibitem [{\citenamefont {Frolov}\ \emph {et~al.}(2020)\citenamefont {Frolov},
  \citenamefont {Manfra},\ and\ \citenamefont {Sau}}]{Frolov_2020}%
  \BibitemOpen
  \bibfield  {author} {\bibinfo {author} {\bibfnamefont {S.~M.}\ \bibnamefont
  {Frolov}}, \bibinfo {author} {\bibfnamefont {M.~J.}\ \bibnamefont {Manfra}},\
  and\ \bibinfo {author} {\bibfnamefont {J.~D.}\ \bibnamefont {Sau}},\
  }\bibfield  {title} {\bibinfo {title} {Topological superconductivity in
  hybrid devices},\ }\href {https://doi.org/10.1038/s41567-020-0925-6}
  {\bibfield  {journal} {\bibinfo  {journal} {Nature Physics}\ }\textbf
  {\bibinfo {volume} {16}},\ \bibinfo {pages} {718} (\bibinfo {year}
  {2020})}\BibitemShut {NoStop}%
\bibitem [{\citenamefont {Wang}\ \emph {et~al.}(2018)\citenamefont {Wang},
  \citenamefont {Kong}, \citenamefont {Fan}, \citenamefont {Chen},
  \citenamefont {Zhu}, \citenamefont {Liu}, \citenamefont {Cao}, \citenamefont
  {Sun}, \citenamefont {Du}, \citenamefont {Schneeloch}, \citenamefont {Zhong},
  \citenamefont {Gu}, \citenamefont {Fu}, \citenamefont {Ding},\ and\
  \citenamefont {Gao}}]{Wang333}%
  \BibitemOpen
  \bibfield  {author} {\bibinfo {author} {\bibfnamefont {D.}~\bibnamefont
  {Wang}}, \bibinfo {author} {\bibfnamefont {L.}~\bibnamefont {Kong}}, \bibinfo
  {author} {\bibfnamefont {P.}~\bibnamefont {Fan}}, \bibinfo {author}
  {\bibfnamefont {H.}~\bibnamefont {Chen}}, \bibinfo {author} {\bibfnamefont
  {S.}~\bibnamefont {Zhu}}, \bibinfo {author} {\bibfnamefont {W.}~\bibnamefont
  {Liu}}, \bibinfo {author} {\bibfnamefont {L.}~\bibnamefont {Cao}}, \bibinfo
  {author} {\bibfnamefont {Y.}~\bibnamefont {Sun}}, \bibinfo {author}
  {\bibfnamefont {S.}~\bibnamefont {Du}}, \bibinfo {author} {\bibfnamefont
  {J.}~\bibnamefont {Schneeloch}}, \bibinfo {author} {\bibfnamefont
  {R.}~\bibnamefont {Zhong}}, \bibinfo {author} {\bibfnamefont
  {G.}~\bibnamefont {Gu}}, \bibinfo {author} {\bibfnamefont {L.}~\bibnamefont
  {Fu}}, \bibinfo {author} {\bibfnamefont {H.}~\bibnamefont {Ding}},\ and\
  \bibinfo {author} {\bibfnamefont {H.-J.}\ \bibnamefont {Gao}},\ }\bibfield
  {title} {\bibinfo {title} {Evidence for majorana bound states in an
  iron-based superconductor},\ }\href {https://doi.org/10.1126/science.aao1797}
  {\bibfield  {journal} {\bibinfo  {journal} {Science}\ }\textbf {\bibinfo
  {volume} {362}},\ \bibinfo {pages} {333} (\bibinfo {year} {2018})},\ \Eprint
  {https://arxiv.org/abs/https://science.sciencemag.org/content/362/6412/333.full.pdf}
  {https://science.sciencemag.org/content/362/6412/333.full.pdf} \BibitemShut
  {NoStop}%
\bibitem [{\citenamefont {Kraus}\ \emph {et~al.}(2012)\citenamefont {Kraus},
  \citenamefont {Diehl}, \citenamefont {Zoller},\ and\ \citenamefont
  {Baranov}}]{Zoller_1Dladder}%
  \BibitemOpen
  \bibfield  {author} {\bibinfo {author} {\bibfnamefont {C.~V.}\ \bibnamefont
  {Kraus}}, \bibinfo {author} {\bibfnamefont {S.}~\bibnamefont {Diehl}},
  \bibinfo {author} {\bibfnamefont {P.}~\bibnamefont {Zoller}},\ and\ \bibinfo
  {author} {\bibfnamefont {M.~A.}\ \bibnamefont {Baranov}},\ }\bibfield
  {title} {\bibinfo {title} {Preparing and probing atomic majorana fermions and
  topological order in optical lattices},\ }\href
  {http://stacks.iop.org/1367-2630/14/i=11/a=113036} {\bibfield  {journal}
  {\bibinfo  {journal} {New Journal of Physics}\ }\textbf {\bibinfo {volume}
  {14}},\ \bibinfo {pages} {113036} (\bibinfo {year} {2012})}\BibitemShut
  {NoStop}%
\end{thebibliography}


\wbe
\appendix

\section{The single-particle energy spectra of mean-field Hamiltonian } 
\label{sec:eigen-mode}

In this appendix, we shall summarize the single-particle energy spectra of mean-field Hamiltonian 
$ H(\bfk)$.
First we fix $\alpha_{\uparrow} = \alpha_{\downarrow}=0$ and shown in Table III.
In particular, we can find the spectra of $\phi_{\uparrow} =\phi_{\downarrow} =0$  and  $\phi_{\uparrow} =\phi_{\downarrow} =\pi/2$  up to a $90$ degree rotation transformation. This is, once we replace $k_x$
 to $k_y$, these two spectra are the same. 


\renewcommand{\arraystretch}{1.5}
 \begin{table*}[htbp]
	\centering  
\begin{tabular}{|c||c|} 
 \hline
$\phi_{\uparrow} =\phi_{\downarrow} =0$  &   
 $E_{\bfk} =   \pm\Big[ \varepsilon_{\bfk }^2  + t_z^2  + 4 |\Delta_p|^2  \left( \sin^2 k_x + \sin^2 k_y \right)$ 
 $+|\Delta_s|^2 \pm 2\sqrt{ \varepsilon_{\bfk }^2 \; t_z^2 +  t_z^2|\Delta_s|^2 +  4 |\Delta_p|^2 |\Delta_s|^2 \sin^2 k_x   }   \;  \Big]^{1/2}$   \\
  \hline
$\phi_{\uparrow} =\phi_{\downarrow} =\pi/2$  &   
 $E_{\bfk} =   \pm\Big[ \varepsilon_{\bfk }^2  + t_z^2  + 4 |\Delta_p|^2  \left( \sin^2 k_x + \sin^2 k_y \right)$ 
 $+|\Delta_s|^2 \pm 2\sqrt{ \varepsilon_{\bfk }^2 \; t_z^2 +  t_z^2|\Delta_s|^2 +  4 |\Delta_p|^2 |\Delta_s|^2   \sin^2 k_y  }   \;   \Big]^{1/2}$   \\
 \hline
   $\phi_{\uparrow} =0, \phi_{\downarrow} =\pi$  &   
 $  E_{\bfk}= t_z \pm  \Big[ \varepsilon_{\bfk}^2    +    4 |\Delta_p|^2 \left(  \sin^2 k_x+  \sin^2 k_y \right)- 4 |\Delta_p|| \Delta_s| \sin k_y +   |\Delta_s|^2      \Big]^{1/2}$ \\
 &
 $  E_{\bfk}= - t_z \pm  \Big[ \varepsilon_{\bfk}^2    +    4 |\Delta_p|^2 \left(  \sin^2 k_x+  \sin^2 k_y \right)+ 4 |\Delta_p| | \Delta_s| \sin k_y +   |\Delta_s|^2      \Big]^{1/2}$
 \\
 \hline
\end{tabular} 
	\caption{ The single-particle energy spectra of mean-field Hamiltonian with  $\alpha_{\uparrow} =\alpha_{\downarrow} =0$.}
\end{table*}

Then we consider  $\alpha_{\uparrow} = 0,  \alpha_{\downarrow}=\pi$  and shown in Table IV.

\renewcommand{\arraystretch}{1.5}
 \begin{table*}[htbp]
	\centering  
\begin{tabular}{ |c||c|}
 \hline
  $\phi_{\uparrow} =\phi_{\downarrow} =\pi/2$    & 
 $ E_{\bfk} =   \pm \left[ \varepsilon_{\bfk }^2  + t_z^2  + 4  |\Delta_p| ^2  \left( \sin^2 k_x   + \sin^2 k_y \right) 
  +|\Delta_s|^2 \pm 2  |t_z|  \sqrt{ \varepsilon_{\bfk }^2  +|\Delta_s|^2 +  4 |\Delta_p|^2  \sin^2 k_y }   \;   \right]^{1/2}$      \\
 \hline
    $\phi_{\uparrow} =0, \phi_{\downarrow} =\pi$   & 
 $ E_{\bfk} =   \pm \left[ \varepsilon_{\bfk }^2  + t_z^2  + 4  |\Delta_p| ^2  \left( \sin^2 k_x   + \sin^2 k_y \right) 
  +|\Delta_s|^2 \pm 2  |t_z|  \sqrt{ \varepsilon_{\bfk }^2  +|\Delta_s|^2 +  4 |\Delta_p|^2  \sin^2 k_x }   \;   \right]^{1/2}$      \\
 \hline
\end{tabular}
	\caption{ The single-particle energy spectra of mean-field Hamiltonian with  $\alpha_{\uparrow} =0, \alpha_{\downarrow} =\pi$.}
\end{table*}

\wee

\end{document}